\documentclass[preprint,12pt]{elsarticle}

\usepackage[T1]{fontenc}
\usepackage[utf8]{inputenc}
\usepackage{lmodern}
\usepackage[nopatch=footnote]{microtype}
\usepackage{graphicx}
\usepackage{booktabs}
\usepackage{multirow}
\usepackage{tabularx}
\usepackage{xltabular}
\usepackage{hhline}
\usepackage[table]{xcolor}
\newcommand{\gridtable}{\footnotesize\arrayrulecolor{black!35}\setlength{\arrayrulewidth}{0.4pt}\setlength{\tabcolsep}{4pt}\renewcommand{\arraystretch}{1.1}}
\usepackage{etoolbox}
\AtBeginEnvironment{table}{\setlength{\belowcaptionskip}{6pt}}
\newcommand{\gridhead}{\rowcolor{black!11}}
\newcommand{\gridzebra}[1]{\rowcolors{#1}{}{black!3.5}}
\newcommand{\genrule}{\hline}
\usepackage{enumitem}
\usepackage{amsmath,amssymb}
\usepackage{tikz}
\usetikzlibrary{positioning,arrows.meta,calc,shapes.geometric,decorations.pathreplacing,fit,backgrounds}
\usepackage{orcidlink}
\usepackage{hyperref}
\usepackage[nameinlink,capitalise,noabbrev]{cleveref}
\usepackage{placeins}
\usepackage[margin=2.5cm]{geometry}
\usepackage{setspace}
\graphicspath{{figures/}}
\definecolor{xblue}{HTML}{2A78D6}
\definecolor{xorange}{HTML}{EB6834}
\definecolor{xgreen}{HTML}{1BAF7A}
\definecolor{xred}{HTML}{C62E2E}
\definecolor{xpurple}{HTML}{7B5EA7}
\definecolor{xgold}{HTML}{B07D12}
\definecolor{xink}{HTML}{2B2B2B}
\definecolor{xmuted}{HTML}{6B6A66}
\newcolumntype{L}[1]{>{\raggedright\arraybackslash}p{#1}}
\newcolumntype{Y}{>{\raggedright\arraybackslash}X}
\IfFileExists{generated/numbers.tex}{

\newcommand{\TwentyPDecThirty}{2000}
\newcommand{\TwentyPObsThirty}{35}
\newcommand{\TwentyPValThirty}{1963}
\newcommand{\TwentyPDiscThirty}{2}

\newcommand{\TwentyPObsTwoForty}{147}
\newcommand{\TwentyPValTwoForty}{1852}
\newcommand{\TwentyPDiscTwoForty}{1}

}{}

\hypersetup{
  hidelinks,
  pdftitle={Execution-Time Validation of Context-Carrying Action Intents in Industrial Cyber-Physical Systems},
  pdfkeywords={Execution-time validation; action intent; edge computing; programmable logic controllers; OPC UA; cyber-physical systems},
  pdfauthor={Luca D'Agati, Giuseppe Tricomi, Fabio Orazio Mirto, Giovanni Merlino},
}

\journal{Journal of Systems Architecture}
\makeatletter
\def\ps@pprintTitle{\let\@oddhead\@empty\let\@evenhead\@empty
  \def\@oddfoot{\footnotesize\itshape Preprint submitted to \@journal\hfill\today}\let\@evenfoot\@oddfoot}
\makeatother

\newcommand{\ioc}[2]{\mathopen{]}#1,#2\mathclose{]}}
\newcommand{\ioo}[2]{\mathopen{]}#1,#2\mathclose{[}}
\usepackage{xspace}
\newcommand{\expPause}{E1\xspace}
\newcommand{\expClasses}{E1-classes\xspace}
\newcommand{\expRos}{E1-ros2\xspace}
\newcommand{\expPolicy}{E1-policy\xspace}
\newcommand{\expCache}{E2\xspace}

\newcommand{\expStore}{E3\xspace}
\newcommand{\expStoreDeadline}{E3-deadline\xspace}
\newcommand{\expStoreAba}{E3-ABA\xspace}
\newcommand{\expStoreStop}{E3-stop\xspace}
\newcommand{\expCtrl}{E4\xspace}
\newcommand{\expCtrlAba}{E4-ABA\xspace}
\newcommand{\expCtrlChurn}{E4-updates\xspace}
\newcommand{\expPlc}{E4-PLC\xspace}
\newcommand{\expChurn}{E5\xspace}
\newcommand{\expTrace}{E5-trace\xspace}
\newcommand{\expChurnScale}{E5-scale\xspace}
\newcommand{\expLog}{E6\xspace}
\newcommand{\expEdge}{E7\xspace}
\newcommand{\expScale}{E8\xspace}
\newcommand{\expNetwork}{E9-impair\xspace}
\newcommand{\expCrash}{E9-crash\xspace}
\newcommand{\expFaults}{E9\xspace}

\begin{document}

\hypersetup{pageanchor=false}% the title page would duplicate the anchor of page 1
\begin{frontmatter}

\title{Execution-Time Validation of Context-Carrying Action Intents in Industrial Cyber-Physical Systems}

\author[unime,cini]{Luca D'Agati\corref{cor1}\orcidlink{0000-0002-9992-4179}}
\ead{ldagati@unime.it}
\author[unime,cini]{Giuseppe Tricomi\orcidlink{0000-0003-3837-8730}}
\author[unime]{Fabio Orazio Mirto\orcidlink{0009-0001-5830-1901}}
\author[unime,cini]{Giovanni Merlino\orcidlink{0000-0002-1469-7860}}
\cortext[cor1]{Corresponding author.}
\affiliation[unime]{organization={Department of Engineering, University of Messina},
            addressline={Contrada Di Dio}, postcode={98166}, city={Messina}, country={Italy}}
\affiliation[cini]{organization={National Interuniversity Consortium for Informatics (CINI)},
            addressline={Via Ariosto 25}, postcode={00185}, city={Rome}, country={Italy}}

\begin{abstract}
In industrial cells, commands from learned policies, operators and schedulers reach the controller after inference, confirmation and transport, and the plant may have changed meanwhile, for instance by an operator pause. Such a command is released, or takes effect, when it is no longer correct. We call this the \emph{execution gap} and ask which component should make the last check, and against which version of which state. The Action Intent Schema (AIS) is a contract in which a producer states how long its decision stays valid and what it assumes. The runtime XAIR rechecks each intent before release, can commit its authorization atomically in the context store, and can release it as a \emph{version-conditional} command, which the controller executes only if its version of the state is unchanged. In controlled ablations with declared ground truth on emulated and physical cells, OpenPLC and ARM and RISC-V boards, a check on a replica always left an unprotected interval: a pause written from a few to over 100\,ms after submission escaped it. Only a check in the scan of a software controller, either a controller-local interlock or the version-conditional command, closed the interval for the state that controller owns; only the version-conditional command also refused commands after a pause that had been undone. A global version revoked up to every valid command under unrelated updates. Versions scoped to what the intent reads, or to the truth of its predicates, did not; with continuous signals on the variables the intent reads, only the latter avoided false revocations, though it burdens the store as predicates accumulate. On one embedded-board core, XAIR validated 148--168 intents/s.
\end{abstract}


\end{frontmatter}

\section*{Abbreviations}
\begingroup\centering\gridtable\gridzebra{2}
\begin{tabularx}{\linewidth}{|l|Y|l|Y|}
\hline
\gridhead\textbf{Abbr.} & \textbf{Meaning} & \textbf{Abbr.} & \textbf{Meaning} \\
\hline
AIS & Action Intent Schema & OPA & Open Policy Agent \\
\hline
CAS & compare-and-set & OPC UA & Open Platform Communications Unified Architecture \\
\hline
CPS & cyber-physical systems & PLC & programmable logic controller \\
\hline
E2E & end-to-end & ROS 2 & Robot Operating System 2 \\
\hline
FPR & false-positive revocation rate & RQ & research question \\
\hline
LAN & local area network & RSS & resident set size \\
\hline
LLM & large language model & SER & stale-release rate \\
\hline
MES & manufacturing execution system & TOCTOU & time of check to time of use \\
\hline
NTP & Network Time Protocol & XAIR & eXecution-time Action Intent Runtime \\
\hline
\end{tabularx}
\par\vspace{2pt}\noindent{\footnotesize\raggedright Testbeds in the evaluation: S single host, D distributed, R reserved cores, P wide-area physical nodes, L local-LAN node, B embedded boards.\par}
\endgroup
\hypersetup{pageanchor=true}

\section{Introduction}
\label{sec:introduction}

Industrial cyber-physical systems (CPS) increasingly combine certified control with machine perception, learned policies, and human supervision~\cite{monostori2016cyber}. Making such learning-enabled systems safe is an open problem~\cite{paoletti2023learningenabled}, and part of it is untouched by faster perception or planning. Inference, queuing, and transport separate the moment an action is decided ($t_d$) from the moment it is released into middleware ($t_m$), and the plant keeps changing in between.

Consider, as an example, a collaborative assembly station in which a task planner built on a language model decides to restart the station's conveyor (stopped while the station waits) and to pick the next part. Inference alone takes about a second, and the proposal may then wait for an operator to confirm it. If, in the meantime, the operator pauses the line to clear a jam, or the manufacturing execution system (MES) switches the station to another product variant, the command that eventually reaches the cell controller is still well formed, but the state it was based on no longer exists. We refer to this failure mode as the \emph{execution gap}, i.e., the release of an action that was admissible when it was proposed but is no longer admissible when it is released.

Several lines of work already check or correct the outputs of such systems, from runtime enforcement and shields to middleware monitors and, more recently, firewalls for planners and gateways for agents built on large language models (LLMs), and since 2025 the same time-of-check-to-time-of-use (TOCTOU) race has been studied for such agents. \Cref{sec:related-work} reviews them. Rechecking an action when it is executed is therefore not new, and we do not claim it as our contribution. What these works leave open is the case of an industrial cell, in which several producers act on a plant state that is spread over a context store, the middleware and the controllers. Each of these components orders updates in its own serialization domain, with its own clock and with replicas of the others' state, so that a check placed in one domain cannot see changes ordered in another. For this reason we regard the problem as an architectural one, in which what the last check can guarantee depends mostly on where it runs and on which version it compares, and much less on how fast it is.

In this paper we address this question with an explicit contract, a runtime that enforces it, and an evaluation of where in the architecture the check has to run. Specifically, we make the following contributions.
\begin{enumerate}[label=(\roman*),leftmargin=*,nosep]
\item We model the execution gap by separating decision, validation, the last check, an optional commit, release and effect, and by distinguishing what holds in the plant from what a check is able to observe. The model shows that any check performed on a replica leaves an unprotected interval, which can only be removed by a check in the serialization domain that owns the context, and that the scope of the compared version determines whether a change undone before the effect is detected at all.
\item We present the Action Intent Schema (AIS), through which each producer declares how long its decision remains valid and which plant conditions it relies on, together with the eXecution-time Action Intent Runtime (XAIR). XAIR adds conditions mandated by the operator and rechecks every intent before it is released, either optimistically or through an atomic commit in the context store. It can also release an intent as a command that the controller owning the context will execute only if its state has not changed in the meantime. This resembles If-Match, compare-and-swap and fencing tokens, but the expected version is stated in the contract by a producer that never talks to the controller, and that is what lets the controller refuse a command built on state it did not see.
\item We evaluate the approach in controlled experiments with ground truth declared before execution, on six testbed configurations (a single host, a container-distributed testbed, reserved server cores, a wide-area node, a local-network node, and an embedded configuration with an ARM and a RISC-V board) and on an OpenPLC runtime, against a policy engine and a controller-local interlock as baselines and with two public industrial datasets. The implementation, the per-trial data and the analysis scripts are publicly available.
\end{enumerate}

Our results are prototype evidence obtained with software controllers, and they do not certify safety: what XAIR enforces is that the assumptions an intent declares, together with those the operator mandates, still hold when it is released, a narrower property than physical safety (\Cref{sec:action-abstraction}). \Cref{sec:related-work} discusses related work, \Cref{sec:execution-gap} positions and formalizes the problem, and \Cref{sec:xair} describes AIS and XAIR, while \Cref{sec:evaluation,sec:discussion} present the evaluation and discuss its implications, among them a placement rule that we turn into a design guide for edge deployments.

\section{Related Work}
\label{sec:related-work}

Terms such as replica, version scope and unprotected interval are defined in \Cref{sec:execution-gap}; for this section, a replica is any copy of the state that may lag its owner.

Since XAIR sits between a decision and its effect, the most obvious comparison is with runtime verification and runtime enforcement. In runtime verification, executions are observed and checked against formal properties, while runtime enforcement can also suppress or modify the outputs that would violate a specification~\cite{bartocci2018monitoring,falcone2011enforcement}. Shields, for example, compute minimally interfering corrections from temporal specifications~\cite{alshiekh2018shield}, ModelPlex synthesizes monitors from verified hybrid-system models~\cite{mitsch2016modelplex}, and command governors reshape continuous inputs so that the system stays within model-based constraint sets~\cite{garone2017governors}. These techniques reason about the output and about the state they observe. A shield placed in front of our assembly station would correct a \texttt{RESUME} that violates its specification, but it would notice that the operator paused the line after the command was proposed only if the line state were part of that specification and were observed again at release, which is exactly the problem we study.

These approaches give guarantees that we do not claim for AIS/XAIR, since they reason about trajectories, controller outputs or the conformance between plant and model. AIS, by contrast, deals with an earlier software boundary, namely a discrete action request whose contextual assumptions may expire before it is released, and its predicates are contracts written by the application rather than invariants derived from a proof. In a real deployment we would therefore expect XAIR to work alongside a shield, a governor or a ModelPlex monitor, without replacing the verified model or the fallback policy on which they depend.
A related line of work deals with runtime assurance. In the Simplex architecture, a decision module hands authority from an advanced controller to a verified one as soon as it detects unacceptable behavior~\cite{seto1998simplex}; SOTER composes such modules for robotics~\cite{desai2019soter}, and safety filters have been proposed to constrain learned controllers~\cite{hsu2024safetyfilter}. All of these architectures decide which controller, or which input, drives the plant. XAIR answers a different question (whether a discrete middleware action should be released at all), and for this reason it can be placed upstream of any of them.

On the middleware side, ROS~2 actions already provide a goal lifecycle with feedback and cancellation, and behavior trees place condition nodes close to the execution of a task~\cite{macenski2022ros2,colledanchise2018bt}. ROSRV and ROSMonitoring go further and intercept ROS messages in order to check formally specified properties~\cite{huang2014rosrv,ferrando2020rosmonitoring}. In both cases, however, what is checked is a message trace against a property of the application. With AIS we take a different approach, in which the validity metadata become part of every producer's request, and the contract also specifies when, and against which version of the context, they are rechecked.

The work closest to ours addresses the same execution boundary, mostly in the context of LLM-based planners. Typed-plan firewalls, for example, mediate the output of a planner as typed intent objects bound to a policy version; they fail closed and produce replayable denial traces before anything reaches CPS control~\cite{petel2026typedplan}. In contract-grounded behavior-tree synthesis, an explicit capability contract of the robot is retrieved before generation, and a validation gate on the robot side is applied before execution~\cite{salfity2026contractbt}. AIS follows the same pattern of typed intents with a contract and a gate. Where it differs is in what is versioned, since we version the context an intent depends on instead of the policy, and in the fact that this context is rechecked at release time, when multiple producers and remote writers may be looking at different versions of the snapshot.

A recent line of work studies the same race for agents built on LLMs. Lilienthal and Hong describe TOCTOU vulnerabilities in agents whose files or API responses change between the check and the use, and they reduce them with prompt rewriting, state-integrity monitoring and tool calls that fuse the check with the use~\cite{lilienthal2025toctou}. Jiang et al. find the same problem in browser-use agents, whose pages change between planning and execution, and validate the page state right before each action~\cite{jiang2026atomicity}. The OpenPort protocol specifies a gateway for the tool access of agents, in which writes are bound to a preflight impact, made idempotent and audited, and an optional state witness fails closed when the state has changed~\cite{zhu2026openport}. Closest to our problem, Shraga et al. study verdict staleness in self-adaptive systems guarded by an LLM, where an approved action may no longer be safe when it is executed, and their Freshness-Bounded Shield estimates how long an approval stays valid from its safety margin and the recent volatility of the features, without a model of the plant~\cite{shraga2026approved}. We share with all of them the view that an approval expires, and the freshness window and the deadline of AIS play the role of their time bounds. The difference lies in where the recheck runs and in what it compares. In these systems the check is made by the agent or by one gateway, against the state that this component can observe, which in the terms of our model is a check on a replica (a state witness, for example, is close to our read-set version but is compared at the gateway). In an industrial cell the state is owned by several components, and we compare the version of exactly what the action reads, or let the controller that owns the state compare its own version in the scan that produces the effect. None of these works compares version scopes or places the last check in the owner of the state, and these two questions are what our evaluation is about.

From a security perspective, the execution gap is an instance of a TOCTOU race. Such races were classically analyzed for file accesses~\cite{bishop1996toctou}, and they have been exploited in database-backed applications whose check-then-act sequences run at weak isolation levels~\cite{warszawski2017acidrain}. Several authorization models deal with attributes that change over time. Usage control makes authorization continuous over mutable attributes, with checks both before and during use~\cite{park2004ucon}, attribute-based access control evaluates policies over subject, object and environment attributes~\cite{hu2014abac}, and general-purpose policy engines such as OPA evaluate these policies at query time over replicated data~\cite{opa2025}. Leases and capabilities with contextual caveats, finally, limit the validity of a grant in time or by context~\cite{gray1989leases,birgisson2014macaroons}. AIS borrows these ideas and applies them at a cyber-physical release boundary, where its preconditions play the role of environment attributes, the read-set version tells whether one of them has changed, and the freshness window and the deadline behave like leases on the decision. In the same way, the optimistic recheck corresponds to the validation phase of optimistic concurrency control~\cite{kung1981optimistic}, the atomic commit to a linearization point for authorization~\cite{herlihy1990linearizability}, and the actuation log to an append-only event log, as used in log-based event processing~\cite{kleppmann2019online}. None of these patterns is new, and we do not claim otherwise; what we add is a precise account of where each of them applies between producers, context and actuators, together with a measurement of what each one leaves open.

Timing analysis of embedded cause-effect chains bounds how old the data behind an actuation can be, across automotive multi-rate task chains and ROS~2 processing chains~\cite{becker2017causeeffect,tang2024ros2chains}. Such bounds answer how stale an input may be when it reaches the actuator, not whether the assumption an action was based on still holds, which can fail after a fresh input, as when an operator pauses a line one millisecond after the decision. AIS uses elapsed time as one condition, the freshness window and the deadline, and adds the contextual predicates and their versions.

At the other end, a PLC program routinely interlocks a command on the state it owns, in the scan that executes it~\cite{john2010iec61131}, and redundant controllers checkpoint and transfer that state so that it has one owner at a time~\cite{johansson2026controllerredundancy}. Conditional writes generalize the pattern beyond controllers: an HTTP request with \texttt{If-Match} is applied only if the resource still carries the entity tag the client observed~\cite{rfc9110}, a compare-and-set (CAS) primitive. The same pattern is familiar in distributed systems: a fencing token lets a resource reject a request that carries a token older than the latest it has seen~\cite{kleppmann2017ddia}, the Kubernetes API applies an update only if the version of the object that the client read is still the current one~\cite{k8sapi}, and a Redis transaction on watched keys is aborted if any of them changed in the meantime~\cite{redistx}. In all of them the owner of the resource compares a version it controls. The version-conditional command of \Cref{sec:execution-gap} is such a write at the controller, and the controller-local interlock is the baseline it must improve on. Both check the state where it is owned and differ only in whether they detect a change undone before the effect (\Cref{sec:evaluation}). What AIS adds is the contract that tells the controller which version to expect, carried from a producer that never talks to the controller.

Interoperable information and control flows between industrial components are supported by Industry~4.0 architectures, by OPC~UA and by IEC~61499~\cite{monostori2016cyber,mahnke2009opcua,vyatkin2011iec}, but to our knowledge none of these mechanisms decides whether a well-formed action is still admissible at the moment it is released. AIS is meant to travel over these mechanisms and to give all their producers a common admissibility contract. It does not try to standardize what the context means, and only defines whether an action is still admissible with respect to that context when it is released.

Digital twins keep synchronized virtual counterparts of physical assets~\cite{tao2019digitaltwin}, and in real-time databases the absolute and relative temporal consistency of changing data has been defined formally~\cite{ramamritham2004rtdb}. The freshness window and the read-set of AIS are close to these notions, with the difference that in AIS such conditions are attached to the authorization decision of a single producer and are evaluated before release.

This separation between proposing and authorizing an action becomes even more important with learning-enabled producers. Manufacturing already shows the trend, for example in multi-agent task planning with LLMs for human--robot collaborative assembly~\cite{wang2025llmassembly}. In all these systems the output of the model is still only a proposal, because inference alone cannot establish that a supervisory pause, the state of a resource or a deadline has not changed by the time the command reaches the middleware. For this reason we place XAIR downstream of such models, as an authorization boundary that does not do any planning itself.

We summarize the comparison in \Cref{tab:positioning}. As the table shows, we see AIS/XAIR neither as the first execution-time monitor or gate nor as the first mechanism that revalidates an action before it runs. In our view the contribution lies elsewhere, in five points that we did not find together in previous work: (i) the specialization to industrial cells, where several producers act on state that controllers own, (ii) a systematic comparison of the serialization domains in which the last check can run, (iii) the distinction between global, read-set and predicate-truth versions, (iv) a version-conditional command whose last check runs in the controller that owns the state, and (v) an empirical protocol that measures stale releases, false revocations, latency and placement together. The protocol varies the context-consistency class and keeps the payloads and the drift ground truth fixed, apart from \expLog, which replays a production log and submits every decision through every gate.

\begin{table}[!t]
\caption{Positioning of AIS/XAIR relative to adjacent runtime mechanisms.}
\label{tab:positioning}
\centering
\gridtable
\gridzebra{2}
\begin{tabularx}{\linewidth}{|L{3.1cm}|L{3.0cm}|L{2.8cm}|Y|}
\hline
\gridhead\textbf{Approach} & \textbf{Runtime abstraction} & \textbf{Intervention} & \textbf{Distinction from AIS/XAIR} \\
\hline
Runtime enforcement, shields & Formal trace or temporal property & Suppress or correct output & No common producer intent or context version \\
\hline
ModelPlex, governors & Verified model, admissible set & Validate or reshape input & Requires plant and model assumptions \\
\hline
Simplex, SOTER & Advanced and fallback controllers & Switch execution authority & Controller-level, not intent-level, assurance \\
\hline
ROSRV, ROSMonitoring & ROS messages and traces & Intercept or observe events & Properties external to intent and context version \\
\hline
Policy engines (OPA) & Policies over replicated data & Allow or deny per query & Versions and release audit left to the deployment \\
\hline
UCON, ABAC, leases & Mutable attributes, timed grants & Continuous or expiring authorization & No read-set version at a cyber-physical release \\
\hline
Typed plans, contract behavior trees & Typed intent or capability contract & Normalize or reject a plan & Static contract; no recheck of changing context \\
\hline
TOCTOU defenses for LLM agents & Tool calls, files, web pages & Revalidate state before use & One agent's view; no version scope or owner-side check \\
\hline
Agent tool gateways (OpenPort) & Governed tool call, state witness & Fail closed on changed state & Witness compared at the gateway, i.e., on a replica \\
\hline
Verdict freshness (FBS) & LLM-approved adaptation action & Bound the lifetime of an approval & Time bound from margin and volatility; no versions \\
\hline
Input-delay analysis, PLC interlocks & Cause-effect chains; owned state & Bound how old the input is; refuse in the scan & Elapsed time or local state only; no producer contract or version \\
\hline
Conditional writes (If-Match, CAS, fencing tokens) & Entity tag of an owned resource & Apply or reject the write & Owner's version only; no producer contract or deadline \\
\hline
\textbf{AIS/XAIR} & Context-carrying multi-producer intent & Validate at $t_v$; recheck, commit, or version-conditional command & Read-set or predicate-truth versions; release and effect audited \\
\hline
\end{tabularx}
\end{table}

\section{The Execution Gap}
\label{sec:execution-gap}

In industrial automation, functions are usually layered by timescale and by responsibility. Certified programmable logic controller (PLC) programs implement the hard real-time interlocks, whereas MES, supervisory interfaces, perception and learned policies operate on top of them~\cite{monostori2016cyber}. In the upper layers we observe three trends that widen the interval between a decision and the corresponding actuation. First, producers are becoming more diverse (vision pipelines, learned policies, extended-reality clients), and each of them encodes validity in its own way, if at all. Second, components are increasingly decoupled through queues and remote calls. Third, operators, MES updates and autonomous policies keep changing the plant while intents are still in flight. For this reason we place the gap in an \emph{execution-governance} layer, which works at timescales between milliseconds and seconds (i.e., between sub-50\,ms interlocks and minute-scale planning) and in which a discrete, non-reflex command such as restarting a conveyor is authorized for release into middleware. XAIR is limited to this layer, and deterministic control, transport and interoperability are still provided by PLC, Robot Operating System~2 (ROS~2) and Open Platform Communications Unified Architecture (OPC~UA) components~\cite{macenski2022ros2,mahnke2009opcua,vyatkin2011iec}. Since we intend the runtime to run on an edge node next to the controller, we also consider its cost on such a node as part of the design problem.

Throughout the paper we return to the assembly station of \Cref{sec:introduction}. Its line has a run mode, \texttt{line.state}, which the operator sets to \texttt{PAUSED}, for instance to clear a jam, and back to \texttt{RUN}. A planner decides at time $t_d$ to restart the station's conveyor (\texttt{RESUME}), assuming \mbox{\texttt{line.state == 'RUN'}} and \mbox{\texttt{gripper.state == 'OPEN'}}. The effect of \texttt{RESUME} is a running conveyor, and a pause stops it. The cell PLC owns the line state, and a context store holds a replica of it. If the operator pauses the line before the command is released, the release is stale, and its effect would start the conveyor on a paused line. If the operator pauses and resumes the line in between, the state is correct again at release, but the line stopped in the meantime, and a policy may well forbid the restart.

The execution gap is a late-binding authorization problem. The model below suggests one design principle: \emph{run the last check where the state it reads is owned, in its serialization domain, and compare a version that covers only what the action depends on}. The rest of the section defines the instants, the versions and the places where the last check can run.

\subsection{Intents, instants, and admissibility}
We first fix what an intent contains, when its checks happen and when it is admissible. Let an intent be ${I=(t_d,w,d,P,S,\alpha,r,p)}$, where $t_d$ is the decision time, $w$ a freshness window, $d$ an optional deadline, $P$ contextual preconditions, $S$ hard constraints, $\alpha$ the action descriptor, $r$ the target resource, and $p$ a scheduling priority. The two durations have different roles: $w$ bounds the time elapsed since the decision when it is validated, and $d$ bounds it when it is released, with $d=w$ when no deadline is declared. The producer declares both $P$ and $S$; \Cref{sec:action-abstraction} adds operator-mandated predicates, and \Cref{tab:notation} summarizes the notation. Let $C(t)$ be the context snapshot in the store at time~$t$. Where a controller owns part of the context, such as the line state, let $X(t)$ be the controller's own state, of which $C$ holds a lagging replica.

\begin{table}[tbp]
\caption{Notation of the execution-gap model.}
\label{tab:notation}
\centering
\gridtable
\gridzebra{2}
\begin{tabularx}{\linewidth}{|l|Y|}
\hline
\gridhead\textbf{Symbol} & \textbf{Meaning} \\
\hline
$I=(t_d,w,d,P,S,\alpha,r,p)$ & intent: decision time, freshness window, deadline, preconditions, hard constraints, action, target, priority \\
\hline
$C(t)$, $X(t)$ & context snapshot in the store; state owned by a controller, replicated in $C$ \\
\hline
$t_v$, $t_g$, $t_c$, $t_m$, $t_a$ & validation, gate read (optimistic), commit (atomic), middleware release, effect \\
\hline
$t_\ell$ & instant of the last check, $t_g$ or $t_c$ (\Cref{eq:order}) \\
\hline
$\Delta_g$, $\Delta_r$ & $t_\ell-t_v$; $t_m-t_\ell$, the interval no store check covers \\
\hline
$\ioc{a}{b}$, $\ioo{a}{b}$ & European interval notation: a bracket facing outward excludes its endpoint, so $\ioc{a}{b}=\{x:a<x\le b\}$ and $\ioo{a}{b}=\{x:a<x<b\}$ \\
\hline
$a$, $\hat a$, $\varepsilon$ & true and measured time elapsed since the decision; bound on the clock offset \\
\hline
$A_v$, $A$ & admissibility for validation and for release (ground truth) \\
\hline
$V_v$, $V_g$ & admissibility as validated and as rechecked at the gate \\
\hline
$R(I)$, $\nu_R$, $\pi_I$ & read-set; its latest value-change version; latest truth-change version of $I$'s predicates \\
\hline
$Y(I)$ & release indicator; a stale release is $Y=1\land\neg A(I,t_m)$ \\
\hline
SER, FPR & stale-release rate: share of obsolete intents that are released; false-positive revocation rate: share of valid intents that are revoked \\
\hline
\end{tabularx}
\end{table}

\begin{figure}[tbp]
  \centering
  \resizebox{\linewidth}{!}{%
  \begin{tikzpicture}[x=1cm,y=0.43cm,font=\scriptsize,
    actor/.style={draw=#1!75!black,fill=#1!10,rounded corners=3pt,line width=0.7pt,minimum width=2.2cm,minimum height=0.62cm,font=\footnotesize\bfseries,text=xink},
    msg/.style={-{Stealth[length=1.8mm,width=1.4mm]},line width=0.6pt,xink!80},
    rel/.style={-{Stealth[length=2.2mm,width=1.7mm]},line width=1.3pt,#1},
    lab/.style={font=\scriptsize,inner sep=1.2pt,text=xink!85},
    inst/.style={circle,fill=#1,inner sep=1.6pt},
    ilab/.style={font=\scriptsize\bfseries,text=#1,inner sep=1pt},
    band/.style={fill=#1,draw=none,rounded corners=4pt},
    phase/.style={font=\scriptsize\bfseries,text=#1,anchor=north west,inner sep=2pt},
    gap/.style={fill=xred!80,draw=none},
    gaplab/.style={font=\scriptsize,text=xred!85!black,align=left,anchor=west,inner sep=1pt}]
    % phase bands
    \path[band=black!4]   (-0.35,-0.3)  rectangle (14.55,-4.6);
    \path[band=xblue!7]   (-0.35,-5.0)  rectangle (14.55,-10.3);
    \path[band=xorange!9] (-0.35,-10.7) rectangle (14.55,-17.75);
    \path[band=xgreen!10] (-0.35,-18.15) rectangle (14.55,-23.2);
    \node[phase=xmuted]           at (-0.3,-0.35)  {validation};
    \node[phase=xblue!80!black]   at (-0.3,-5.05)  {optimistic recheck};
    \node[phase=xorange!85!black] at (-0.3,-10.75) {atomic commit};
    \node[phase=xgreen!60!black]  at (-0.3,-18.2) {conditional command};
    % actors and lifelines
    \foreach \x/\c/\n in {0.9/xmuted/Producer, 4.25/xorange/XAIR, 7.6/xblue/Store, 10.95/xblue/Gateway, 14.2/xgreen/{Controller}}{
      \draw[black!30,dash pattern=on 2pt off 2pt] (\x,0.3)--(\x,-23.2);
      \node[actor=\c] at (\x,1.0) {\n};}
    % --- validation
    \node[inst=xink] at (0.9,-1.4) {}; \node[ilab=xink,below] at (0.9,-1.55) {$t_d$};
    \draw[msg] (0.9,-1.4)--node[lab,above]{AIS intent}(4.25,-1.4);
    \draw[msg] (4.25,-2.3)--node[lab,above]{read $C,\ \nu_R$}(7.6,-2.3);
    \draw[msg] (7.6,-3.1)--(4.25,-3.1);
    \node[inst=xorange] at (4.25,-3.1) {}; \node[ilab=xorange!85!black,left] at (4.15,-3.1) {$t_v$};
    \draw[msg] (4.25,-3.95)--node[lab,above,pos=0.76]{\textsc{Execute}, $\nu_R(t_v)$}(10.95,-3.95);
    % --- optimistic recheck
    \draw[msg] (10.95,-6.1)--node[lab,above,pos=0.22]{recheck}(4.25,-6.1);
    \draw[msg] (4.25,-6.8)--node[lab,above]{read}(7.6,-6.8);
    \draw[msg] (7.6,-7.8)--node[lab,below]{$C,\ \nu_R$}(4.25,-7.8);
    \node[inst=xblue] at (7.6,-7.8) {}; \node[ilab=xblue!80!black,right] at (7.7,-7.8) {$t_g$};
    \draw[msg] (4.25,-8.8)--node[lab,above,pos=0.82]{unchanged}(10.95,-8.8);
    \fill[gap] (10.83,-7.8) rectangle (11.07,-9.6);
    \node[gaplab] at (11.2,-8.25) {unprotected $\ioc{t_g}{t_m}$};
    \draw[rel=xred] (10.95,-9.6)--node[lab,below,pos=0.62,text=xred!85!black]{release}(14.2,-9.6);
    \node[inst=xred] at (10.95,-9.6) {}; \node[ilab=xred,left] at (10.8,-9.6) {$t_m$};
    % --- atomic commit
    \draw[msg] (10.95,-11.55)--node[lab,above,pos=0.22]{commit}(4.25,-11.55);
    \draw[msg] (4.25,-13.45)--node[lab,above]{one atomic step}(7.6,-13.45);
    \node[lab,anchor=north east] at (7.45,-13.6) {$\nu_R$, elapsed time, log};
    \node[draw=xorange!80!black,fill=xorange!25,rounded corners=1.5pt,minimum width=0.36cm,minimum height=0.55cm] at (7.6,-14.2) {};
    \node[inst=xorange] at (7.6,-14.2) {}; \node[ilab=xorange!85!black,right] at (7.85,-14.2) {$t_c$};
    \draw[msg] (7.6,-14.95)--(4.25,-14.95);
    \draw[msg] (4.25,-15.75)--node[lab,above,pos=0.76]{committed}(10.95,-15.75);
    \fill[gap] (10.83,-14.2) rectangle (11.07,-16.45);
    \node[gaplab] at (11.2,-14.75) {unprotected $\ioc{t_c}{t_m}$};
    \draw[rel=xred] (10.95,-16.45)--node[lab,above,pos=0.62,text=xred!85!black]{release}(14.2,-16.45);
    \node[inst=xred] at (10.95,-16.45) {}; \node[ilab=xred,left] at (10.8,-16.45) {$t_m$};
    \draw[msg,densely dashed,xmuted] (7.6,-17.15)--node[lab,below,pos=0.3,text=xmuted]{or: log consumer at the actuator}(14.2,-17.15);
    % --- conditional command (fixed layout)
    \draw[rel=xgreen!60!black] (10.95,-20.2)--
      node[lab,above=1pt,align=center,text=xgreen!40!black]{execute if version\\and deadline hold}(14.0,-20.2);
    \node[draw=xgreen!60!black,fill=xgreen!25,rounded corners=1.5pt,minimum width=0.36cm,minimum height=0.75cm] at (14.2,-20.2) {};
    \node[inst=xgreen!60!black] at (14.2,-20.2) {};
    \node[ilab=xgreen!50!black,anchor=south] at (14.2,-19.45) {$t_a$};
    \node[font=\scriptsize,text=xgreen!40!black,align=right,anchor=north east,inner sep=1pt] at (13.8,-20.6) {one scan checks and acts:\\ no unprotected interval};
    \draw[msg] (14.0,-22.4)--node[lab,below]{executed or refused}(10.95,-22.4);
  \end{tikzpicture}}
  \caption{Validation, the two release modes, and the version-conditional command that can follow either. Changes on the read-set committed before the gate's store read ($t_g$) or the commit ($t_c$) block release; red bars mark the intervals in which a change is not observed. The version-conditional command checks the context where it is owned, in the controller scan that produces the effect ($t_a$); a controller-local interlock takes the same place but checks the predicates instead of the version.}
  \label{fig:execution-timeline}
\end{figure}
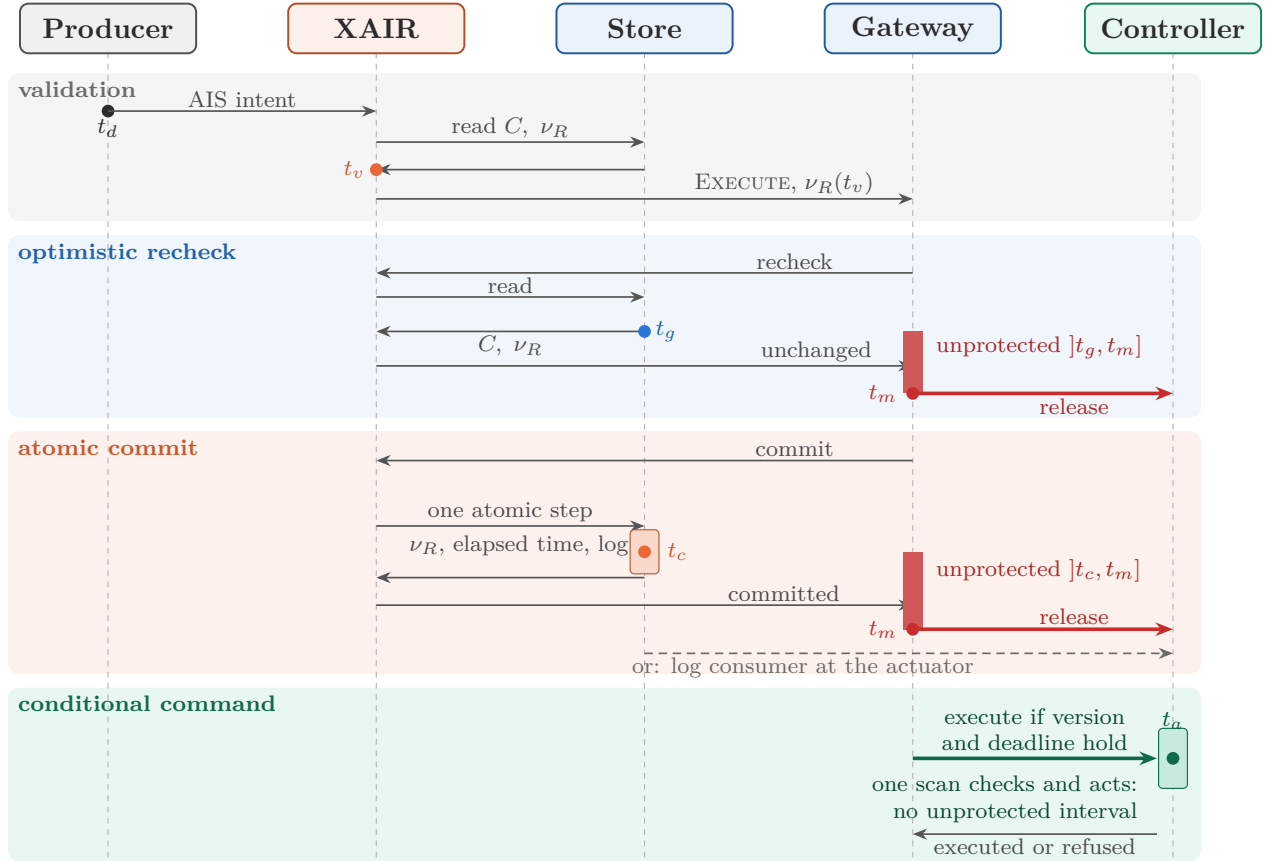

We distinguish the instants of \Cref{fig:execution-timeline}: decision $t_d$, validation $t_v$, the last observation of the store before release, the middleware release $t_m$, and the actuator's effect $t_a$. Two release modes, detailed in \Cref{sec:xair}, place the last observation differently. In the \emph{optimistic} mode the actuator gateway rechecks the intent against the store at $t_g$, a recheck we call the \emph{gate}. In the \emph{atomic} mode the store checks and records the authorization in one step at $t_c$, the \emph{commit}. Writing $t_\ell$ for the instant of the last check, which is $t_g$ in the optimistic mode and $t_c$ in the atomic one, the instants are ordered as
\begin{equation}
\label{eq:order}
t_d<t_v\le t_\ell<t_m\le t_a,\qquad t_\ell=\begin{cases}t_g & \text{optimistic mode,}\\ t_c & \text{atomic mode,}\end{cases}
\end{equation}
and we let ${\Delta_g=t_\ell-t_v}$ and ${\Delta_r=t_m-t_\ell}$. The gate never observes $C(t_m)$, because the release follows its last read; keeping $t_g$ and $t_m$ apart in the model is what makes the residual risk measurable.

Let ${a(t)=t-t_d}$ be the time elapsed since the decision. An intent is \emph{admissible for validation} at $t$, written $A_v(I,t)$, if ${a(t)\le\min(w,d)}$, $P(C(t))$, and $S(I,C(t))$ hold. It is \emph{admissible for release}, $A(I,t)$, under the same conditions with ${a(t)\le d}$ as the time condition. Both define ground truth; what a guard actually decides is captured by $V_v$ and $V_g$ below.

In practice the producer and the checking process often run on different machines, so the checker can only compute the elapsed time from its own clock and the producer's $t_d$. The measured elapsed time $\hat a$ is then off by the offset between the two clocks. If we can assume that this offset never exceeds a declared bound $\varepsilon$, it is enough for the check to accept ${\hat a\le d-\varepsilon}$ in order to guarantee ${a\le d}$. The bound itself cannot be verified by the checker; the atomic commit, for instance, refuses timestamps that are more than $\varepsilon$ ahead of its clock, but this only catches some of the cases in which the assumption is violated.

\subsection{Versions and what a check observes}
\begin{figure}[!t]
  \centering
  \resizebox{\linewidth}{!}{%
  \begin{tikzpicture}[x=1cm,y=1cm,font=\scriptsize,
    ev/.style={circle,fill=#1,inner sep=1.6pt},
    adv/.style={rectangle,fill=#1,minimum width=0.22cm,minimum height=0.22cm,inner sep=0pt},
    rowlab/.style={anchor=east,font=\scriptsize\bfseries,text=xink},
    out/.style={font=\scriptsize,align=center,text=#1},
    hdr/.style={font=\scriptsize\bfseries,align=center,text=xink}]
    % event axis
    \draw[-{Stealth[length=2mm]},semithick,xink] (0.4,0)--(10.2,0) node[above left,font=\scriptsize]{time};
    \node[ev=xorange] at (0.8,0) {}; \node[below=2pt,font=\scriptsize\bfseries,text=xorange!85!black] at (0.8,0) {$t_v$};
    \node[ev=xblue] at (9.6,0) {}; \node[below=2pt,font=\scriptsize\bfseries,text=xblue!80!black] at (9.6,0) {$t_g$};
    \foreach \x/\l in {2.6/{unrelated sensor\\update},4.5/{pressure\\$120\to180$},6.5/{line\\paused},8.1/{line\\resumed}}{
      \draw[xmuted,densely dotted] (\x,0.1)--(\x,-3.75);
      \node[ev=xink] at (\x,0) {};
      \node[above=3pt,align=center,font=\scriptsize] at (\x,0) {\l};}
    \draw[decorate,decoration={brace,amplitude=4pt},xred] (6.2,1.2)--(8.4,1.2) node[midway,above=4pt,font=\scriptsize\bfseries,text=xred]{A--B--A};
    % version rows
    \foreach \y/\l in {-0.9/{global version},-1.65/{read-set version $\nu_R$},-2.4/{predicate-truth version $\pi_I$},-3.15/{no version (predicate only)}}{
      \node[rowlab] at (0.55,\y) {\l};
      \draw[black!25,line width=2.2pt] (0.8,\y)--(9.6,\y);}
    \foreach \x in {2.6,4.5,6.5,8.1} \node[adv=xmuted] at (\x,-0.9) {};
    \foreach \x in {4.5,6.5,8.1} \node[adv=xmuted] at (\x,-1.65) {};
    \foreach \x in {6.5,8.1} \node[adv=xmuted] at (\x,-2.4) {};
    % gate decisions
    \node[hdr] at (13.0,1.25) {gate decision at $t_g$};
    \draw[xmuted] (11.3,1.0)--(14.7,1.0);
    \node[hdr] at (11.9,0.55) {no pause};
    \node[hdr] at (14.1,0.55) {pause--resume};
    \node[out=xgold!85!black] at (11.9,-0.9) {revoke (false)};
    \node[out=xgold!85!black] at (11.9,-1.65) {revoke (false)};
    \node[out=xgreen!50!black] at (11.9,-2.4) {release};
    \node[out=xgreen!50!black] at (11.9,-3.15) {release};
    \node[out=xgreen!50!black] at (14.1,-0.9) {revoke};
    \node[out=xgreen!50!black] at (14.1,-1.65) {revoke};
    \node[out=xgreen!50!black] at (14.1,-2.4) {revoke};
    \node[out=xred] at (14.1,-3.15) {release: history\\ violation};
    \node[adv=xmuted,label={[font=\scriptsize]right:version advances}] at (0.9,-4.15) {};
  \end{tikzpicture}}
  \caption{Which updates advance each version, for an intent whose predicates are \mbox{\texttt{line.state == 'RUN'}} and \mbox{\texttt{press.p < 250}}. Between validation ($t_v$) and the gate ($t_g$) an unrelated sensor reports and the pressure changes but stays below its threshold; in the second case the line is also paused and resumed. A global or read-set version advances and revokes the valid intent without the pause, a predicate-truth version advances only when a predicate changes truth value, and a gate without a version releases even after the pause.}
  \label{fig:versions}
\end{figure}
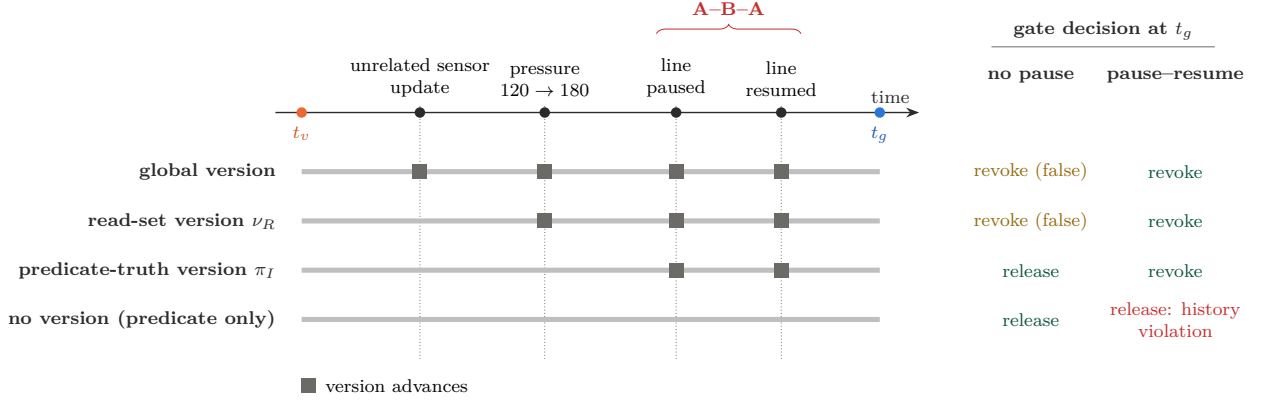

Rechecking the predicates at the gate only tells us whether they hold now. To know whether the context has changed since validation we also need a version, and this subsection defines the three scopes that a version can have. Let $R(I)$ be the \emph{read-set} of $I$, i.e., the context paths referenced by its predicates; for \texttt{RESUME}, for instance, ${R=\{\texttt{line.state},\texttt{gripper.state}\}}$. We denote by $\nu_R(t)$ the latest version at which any path in $R$ (or an ancestor or descendant of one) changed value by time $t$. Since versions are computed on the stored values before the canonicalization of \Cref{sec:predicate-grammar}, we also treat a change of type or representation, such as $1$ to \texttt{true}, as a change of value. With the predicate language of \Cref{sec:predicate-grammar}, $P$ and $S$ depend on the context only through these paths. Validation-time admissibility is then
\begin{equation}
\label{eq:validity-v}
V_v(I)=A_v(I,t_v),
\end{equation}
and gate admissibility is
\begin{equation}
\label{eq:validity-p}
V_g(I)=V_v(I)\land A(I,t_g)\land \nu_R(t_g)=\nu_R(t_v).
\end{equation}
The two conditions also differ in the elapsed time they allow. Validation compares the elapsed time with $w$, whereas the gate compares it with $d-\varepsilon$, and not with $\min(w,d)$, so that a decision that was fresh at $t_v$ can still be too old at $t_g$.
With the version condition the gate is conservative with respect to any value change on $R$ in $\ioc{t_v}{t_g}$, even one that is undone before $t_g$, whereas updates outside $R$, and rewrites that leave a value unchanged, are ignored. A \emph{global} version, which is advanced by every accepted update, corresponds to the special case in which $R$ contains all paths; with such a version, \texttt{RESUME} would be revoked whenever any sensor in the cell reports a new value. At the other extreme, one can version the predicates themselves. Let $\pi_I(t)$ be the latest version at which any predicate of $I$ was registered or changed its truth value. If we replace $\nu_R$ with $\pi_I$ in \Cref{eq:validity-p}, value changes that keep every predicate true (for example, a pressure that changes but remains below its threshold) no longer revoke the intent. A predicate that became false and then true again before $t_g$ is still blocked, provided that each intermediate state was written to the store. Note that versions are only meaningful if every accepted update advances them atomically with the snapshot. \Cref{fig:versions} illustrates, for one sequence of updates, which versions advance and how each gate decides.

\subsection{What a check can still get wrong}
If ${Y(I)=1}$ denotes release, a \emph{stale release} is ${Y(I)=1\land\neg A(I,t_m)}$, a command that crossed into middleware while not admissible, and it is the first of three failures that a gate can still allow after passing $V_g$. Consider a gate that tests $\nu_R$, $P$, and $S$ at $t_g$ and accepts ${\hat a\le d-\varepsilon}$. If the clock offset is within $\varepsilon$, it releases only when ${V_g=1}$, and so rules out every stale release caused by a change committed before $t_g$. Two cases remain: a change committed in $\ioc{t_g}{t_m}$ that falsifies $P$ or $S$, and an elapsed time that crosses $d$ in that interval, which is why the deadline is guaranteed at the check and, at $t_m$, only if a margin or a release guard covers the interval.

The second failure concerns controller-owned context, where the relevant state at the effect is the controller's. A \emph{stale effect} is an effect at $t_a$ with $P$ or $S$ false when evaluated with $X(t_a)$ in place of its replica in $C$. Because $C$ lags $X$, a release that is admissible on $C(t_m)$ can still have a stale effect.

The third failure is not pointwise. Admissibility does not require a predicate to have remained true between validation and release, and a \emph{history violation} is then a release with $A(I,t_m)$ although ${P(C(\tau))\land S(I,C(\tau))}$ was false for some ${\tau\in\ioo{t_v}{t_m}}$: the A--B--A case, as when the operator pauses and resumes the line in between. It is not stale, yet it violates a policy that requires uninterrupted validity. On controller-owned context the same holds with $X$ in place of $C$, between the state the check observed and $t_a$. Predicate-truth versions enforce uninterrupted validity up to the check, read-set versions more conservatively, and a \emph{predicate-only} gate, which re-evaluates $P$ and $S$ at $t_g$ without a version, not at all. Which of these a deployment wants is a policy decision, configured at the gate, and we report history violations separately.

\subsection{Where the last check can run}
With the atomic commit, the last check is moved into the store. We denote by $t_c$ the commit of an authorization record, which the store accepts only if ${\nu_R(t_c)=\nu_R(t_v)}$ and ${\hat a(t_c)\le d-\varepsilon}$ on the store clock, within a single operation serialized with all context updates. Any update that changes a value on $R$ is then ordered either before the commit (in which case the commit is refused) or after it. Hence, whenever the offset between the store and producer clocks is within $\varepsilon$, a committed record implies $A(I,t_c)$. We note that this gives a linearizable authorization~\cite{herlihy1990linearizability} on a single store, but it does not make the actuation atomic. The command still reaches the middleware at ${t_m>t_c}$, and the commit cannot prevent a change in $\ioc{t_c}{t_m}$, which we call a post-commit invalidation. The commit does not shorten the unprotected interval; what it provides is an exact ordering, since every committed authorization was admissible at commit time and every update is ordered with respect to it. As a result, post-commit invalidations can be detected from the store versions, and the committed records form a log from which an idempotent actuator can apply each command at most once.

Removing the unprotected part of $\ioc{t_g}{t_a}$ or $\ioc{t_c}{t_a}$ requires the actuator itself to condition its effect on the context, within one serialization domain. This is possible where the controller in front of the actuator owns the context that matters, as the cell PLC owns the line state. The controller can then execute the command only if its own version of that state still equals the one the check observed and the deadline has not passed, in the scan that produces the effect. We call such a command \emph{version-conditional}. Where the controller cannot check absolute time, the deadline is enforced at the gateway's check or the commit instead (as on the PLC runtime of \Cref{sec:evaluation}). This rules out stale effects on controller-owned context under five assumptions:
\begin{enumerate}[label=(A\arabic*),leftmargin=*,nosep]
\item the controller is the only writer of that state;
\item its version advances on every change of a variable that $P$ or $S$ reads, and ideally on no other, because a version that covers more state revokes valid commands (\Cref{sec:evaluation});
\item the replica in $C$ carries each state together with its version, so the version the check observed belongs to the state it evaluated;
\item a version value is never reused while an intent can still carry it, including across restarts;
\item check and effect happen in one scan, so an effect means the start of the commanded action.
\end{enumerate}

Under these assumptions the absence of stale effects follows by construction, as for any compare-and-set. For context owned elsewhere, such as an MES order, the controller holds only a replica and the interval remains. An intent may also depend on several owners at once, for instance on the line state at a controller and on a recipe in the MES. Its read-set then splits by owner into $R_1,\dots,R_k$. One controller executes the command and can compare a version only for the part it owns, so the other parts remain checks on a replica at $t_g$ or $t_c$, with their intervals. The guarantee therefore holds per subset, because no single serialization domain closes every unprotected interval short of a distributed transaction, which we do not provide. The other guarantees of the paper also hold only under assumptions of this kind, and \Cref{tab:guarantees} lists them all, with what happens when one of them is violated.

The natural alternative is a \emph{controller-local interlock}, which executes the command only if $P$ and $S$ hold with $X(t_a)$ in that scan, also rules out stale effects, but not history violations on the controller's state. If the line is paused and resumed after the check observed it and before the scan, the interlock sees a running line and executes, whereas the version has advanced and the version-conditional command refuses. The two differ as the predicate-only and the versioned gate do, now at the controller.

\subsection{Cost and metrics}
A version condition is fail-closed: it can also revoke valid intents. Let $\lambda$ be the rate of the updates that advance the version compared by the gate: all updates for a global version, value changes on $R(I)$ for a read-set version, and truth flips for predicate-truth versions. If the gate compares versions read $\Delta_g$ apart and updates are independent of submission, a valid intent is revoked with probability
\begin{equation}
\label{eq:fpr}
\Pr[\text{revoke}\mid\text{valid}]=1-e^{-\lambda\Delta_g}\ \text{(Poisson updates)},\qquad \min(1,\lambda\Delta_g)\ \text{(periodic updates)},
\end{equation}
and in both cases the probability is close to $\lambda\Delta_g$, the expected number of updates during $\Delta_g$, as long as that number is small. In practice, then, the cost of a version is the rate admitted by its scope multiplied by the latency of the gate. A narrower scope reduces $\lambda$, while a slower gate (e.g., on an embedded processor or over a wide-area link) increases $\Delta_g$. The same expression also bounds the risk left by the optimistic recheck, if $\lambda_R$, the rate of changes that falsify $P$ or $S$, is used in place of $\lambda$ and $\Delta_r$ in place of $\Delta_g$.

To avoid judging a guard by its own decisions, we fix the ground truth of every experiment before running it. For example, when an experiment pauses the line shortly after a \texttt{RESUME} intent has been built, we know in advance that the intent must not be released. We call $O$ the set of intents that an experiment invalidates in this way before their nominal release (so that ${\neg A}$ holds when they would be released), and $\mathcal{V}$ the set of valid intents, for which $A(I,t)$ holds for all ${t\in[t_v,t_m]}$. Intents with a history violation are kept out of both sets. On these two sets we compute the stale-release rate (SER) and the false-positive revocation rate (FPR),
\begin{equation}
\mathrm{SER}=\frac{\sum_{I\in O}Y(I)}{|O|},\qquad
\mathrm{FPR}=\frac{\sum_{I\in \mathcal{V}}(1-Y(I))}{|\mathcal{V}|},
\end{equation}
which allows us to judge a fail-closed guard both on drift and on legitimate traffic. In the evaluation we give both metrics as counts, and we report stale effects on controller-owned context separately. For these the ground truth is not XAIR's replica $C$ but the controller's own state in the scan that executes the command, $X(t_a)$. A command executed while the line is paused is a stale effect, and one executed on a running line, at a version newer than the one its last check observed, is a history violation. A pause that lands after the scan is therefore harmless, and one that lands between the middleware call and the scan always counts, because the scan then sees it.

An architecture that follows the design principle stated at the start of this section has four obligations. First, validity bounds and preconditions have to travel with the action. Second, the last check has to run in the serialization domain that owns the context, which means in the controller's scan when the controller owns it and otherwise as close to $t_m$ as the transport allows. It should also be anchored to an atomic commit whenever post-commit invalidations must be audited. Third, outcomes must be recorded together with their reasons. Finally, neither malformed intents nor unavailable context may ever authorize a release.

\section{AIS and XAIR}
\label{sec:xair}
\label{sec:reference-model}

\subsection{Action Intents}
\label{sec:action-abstraction}

A middleware command normally specifies what an actuator should do, whereas an \emph{action intent} also states the assumptions under which the command is admissible. Validity thus travels with the payload and can be evaluated at a shared boundary, instead of being implemented as guards that are scattered across the applications. We designed AIS as a logical contract that does not depend on the transport (HTTP, ROS~2, MQTT or OPC~UA); its core fields are listed in \Cref{tab:ais}, and XAIR checks every incoming intent against the complete JSON Schema. When we defined the schema we left out every field that the runtime would not enforce or record, which is also why unknown fields are rejected. Degradation policies follow the same logic: if a producer asked for, say, \texttt{reduce\_speed} as a fallback and the runtime did not implement it, the fallback would silently turn into the original payload, so XAIR accepts only the policies that it is able to apply. In AIS, the running example becomes a \texttt{RESUME} intent that carries the two preconditions of \Cref{sec:execution-gap} and becomes obsolete as soon as an operator pauses the line, and this is also the intent we use in most of the evaluation.

\begin{table}[tbp]
\caption{AIS core fields.}
\label{tab:ais}
\centering
\gridtable
\gridzebra{2}
\begin{tabularx}{\columnwidth}{|l|l|Y|}
\hline
\gridhead\textbf{Field} & \textbf{Type} & \textbf{Role} \\
\hline
\texttt{id} & UUID & Idempotency and audit \\
\hline
\texttt{source} & enum & Producer class (audit) \\
\hline
\texttt{timestamp\_decision} & time & Freshness anchor $t_d$ \\
\hline
\texttt{freshness\_window\_ms} & int & Maximum elapsed time at validation ($w$) \\
\hline
\texttt{deadline\_ms} & int, optional & Maximum elapsed time at release ($d$; default $w$) \\
\hline
\texttt{preconditions} & list & Predicates over $C(t_v)$ \\
\hline
\texttt{safety\_constraints} & list & Producer-declared hard predicates ($S$; not certified) \\
\hline
\texttt{payload.action\_type} & string & Requested action \\
\hline
\texttt{payload.target\_entity} & string & Coordinated resource \\
\hline
\texttt{payload.parameters} & object & Action parameters \\
\hline
\texttt{payload.degradation\_policy} & enum & \texttt{none} or \texttt{reduce\_speed} \\
\hline
\texttt{priority} & int & Conflict ordering \\
\hline
\end{tabularx}
\end{table}

\label{sec:predicate-grammar}
Hard constraints and preconditions are written as conjunctions of comparisons, each of the form $\mathit{Path}\ \mathit{Op}\ \mathit{Literal}$. Here $\mathit{Path}$ is a dot-separated identifier such as \texttt{line.state}, $\mathit{Op}$ is a comparison operator, and $\mathit{Literal}$ is a string, a number or a Boolean. The reason for such a restricted grammar is that the read-set of a conjunction of comparisons can be derived directly from its syntax, and this is what allows the store to maintain read-set and predicate-truth versions on each update. Predicates are evaluated in a typed and deterministic way, without side effects and within a bounded time. Before a comparison, the evaluator applies a canonicalization $\kappa$, which maps the stored value to the type of the literal: a value of that type is kept as it is, a string that parses as that type is converted (as when an MES reports \texttt{"1"} or \texttt{"true"}), and any other value makes the comparison fail closed. Versions, on the other hand, are computed on the stored values before $\kappa$ is applied, so a change of representation advances them even when $\kappa$ maps the old and the new value to the same one, which can revoke a valid intent but can never hide a change. Evaluation fails closed (i.e., the intent is not authorized) if an expression is empty, compares values of incompatible types, exceeds its evaluation time or refers to a context path that does not exist. In principle, disjunctions and bounded temporal operators could be added to the same scheme, provided that every term reads a declared path and that unknown terms still fail closed, but the current prototype does not support them.

We assume that producers are not malicious, although they may be faulty, and that they operate on a controlled network. Under this assumption, it is useful to keep three properties apart because XAIR provides only the first of them in full. The first is \emph{intent validity}, by which we mean that the freshness window, the deadline and every predicate attached to an intent still hold at the last check, under the assumptions collected in \Cref{tab:guarantees}. The second is \emph{policy compliance}, i.e., that every intent carries the predicates the operator requires for its action. AIS validation alone does not give it, since it can check whether the assumptions declared by an intent still hold but not whether the declaration is complete, and an intent that omits the precondition that actually matters is only checked against what it did declare. Operators can close this gap through a server-side policy that maps each action type (matched case-insensitively) to a set of mandatory predicates. Before validation, XAIR appends these predicates to every intent of that type, the gate rechecks them together with the producer's own predicates, and producers have no way to remove them. Such a policy is worth no more than the interface through which it is installed, however, and in the prototype policy changes can therefore be restricted to requests that carry an operator token, which producers do not hold. The third property is \emph{physical safety}, which XAIR does not provide at all: torque limits, collision envelopes and certified stops remain with the safety-rated layer shown at the bottom of \Cref{fig:architecture}.

XAIR records the producer class for auditing, but it authenticates neither producers nor context writers, and we do not address write authorization on the context store. In a deployment, both kinds of clients would have to be authenticated where they enter the governance zone, i.e., on the conduits marked in \Cref{fig:architecture}. The operator interface would also have to be separated from them. We regard this as a prerequisite for deployment and discuss its mapping to IEC~62443 in \Cref{sec:discussion}. It could be implemented without changing the architecture, for example with data-centric trust~\cite{feng2026trust} or with OPC~UA PubSub authentication backed by a Trusted Platform Module (TPM)~\cite{gilles2023opcuapubsubtpm}, but we have implemented neither.

A gate that fails closed is the right default for commands that increase activity, such as \texttt{RESUME}, but it would be the wrong one for a stop, which should not be held back because a precondition has become false or because the store cannot be reached. For this reason the operator's policy also assigns every action type to a class (stop, start, motion, reset, shutdown, or other). The class is never taken from the producer, as AIS has no field for it and the schema rejects unknown fields, and since it follows the action type, a producer can obtain the treatment of a stop only by asking for a stop. XAIR authorizes an intent of the stop class on its schema validity alone. Its declared predicates are kept for the audit without being evaluated. Neither its elapsed time nor a stale or unavailable snapshot can revoke it, and it takes no lock on its target. In a batch it wins over every other intent on the same target, so that no tie-break can revoke it either.

A stop also fences the intents of its target that are already in flight: when it is authorized, XAIR writes its time into the context, and every other intent on that target carries the predicate that this time is not later than the intent's decision time. A \texttt{RESUME} decided before the stop, whether it is validated before or after it, is then falsified at its gate recheck or at its commit, like any other change of what it depends on. We assume that XAIR is the only writer of these paths.

The gateway then releases the stop without a recheck or a commit and reports the release as it does for any other intent. In the atomic mode, this means that a stop is released without a commit, so it does not appear in the actuation log, and the log is not a complete record of the commands released. The policy refuses mandatory predicates for the stop class, whereas it can attach predicates to all actions of another class, for instance, an operator-presence condition to every reset. This path is meant for supervisory stops, as when a planner or an MES halts a station. Emergency stops and the other safety-rated functions keep their own path through the safety PLC or the robot safety controller and never pass through XAIR. These rules are covered by unit tests in the artifact, while the experiments of \Cref{sec:evaluation} use actions that increase activity, apart from the stop that fences them (\expStoreStop).

\subsection{Lifecycle and Release Modes}

\begin{figure}[!t]
  \centering
  \resizebox{\linewidth}{!}{%
  \begin{tikzpicture}[x=0.84cm,y=1.0cm,
    st/.style={draw=#1!75!black,fill=#1!12,rounded corners=2.5pt,minimum width=1.75cm,minimum height=0.6cm,text height=1.6ex,text depth=0.4ex,align=center,font=\scriptsize\scshape,text=xink,semithick},
    st/.default=xblue,
    term/.style={st=#1,double,double distance=1.1pt},
    arr/.style={-{Stealth[length=1.7mm]},semithick,xink},
    main/.style={-{Stealth[length=2.2mm]},line width=1.1pt,xink},
    lab/.style={font=\scriptsize,inner sep=1.5pt,text=xink}]
    % validation (blue)
    \node[st=xorange] (c) at (0,0) {Created};
    \node[st=xorange] (p) at (3.9,0) {Pending};
    \node[st=xorange] (v) at (7.8,0) {Validating};
    % deferred (purple)
    \node[st=xpurple] (dl) at (5.85,1.75) {Delayed};
    \node[st=xpurple] (dg) at (5.85,-1.75) {Degraded};
    % authorized (orange)
    \node[st=xorange] (a) at (12.4,0) {Authorized};
    \node[st=xorange] (cm) at (14.9,-2.7) {Committed};
    % released (green)
    \node[st=xblue] (rel) at (16.6,0) {Released};
    \node[term=xgreen] (ec) at (20.6,1.3) {Effect confirmed};
    % effect problems (gold)
    \node[term=xgold] (f) at (20.6,0) {Failed};
    \node[st=xgold] (uk) at (20.6,-1.3) {Unknown effect};
    % stopped before release (red)
    \node[term=xred] (r) at (16.6,3.4) {Revoked};
    \node[term=xred] (w) at (12.4,-4.1) {Withheld};
    \node[term=xred] (ex) at (5.85,-4.1) {Expired};
    \draw[main] (c)--node[lab,above=2pt]{admit}(p);
    \draw[main] (p)--(v);
    \draw[main] (v)--node[lab,above=2pt]{\textsc{Execute}}(a);
    \draw[main] (a)--node[lab,above=2pt]{released}(rel);
    \draw[arr] (a)--node[lab,above,sloped,pos=0.45]{commit}(cm);
    \draw[arr] (cm)--node[lab,right=1pt,pos=0.45]{released}(rel);
    \draw[arr] (cm.south)--node[lab,right=1pt,pos=0.35]{guard}(w.north east);
    \draw[arr] (a)--node[lab,left=1pt,pos=0.55]{refused}(w);
    \draw[main] (rel)--(ec.west);
    \draw[arr] (rel)--(f);
    \draw[arr] (rel)--(uk.west);
    \draw[arr,densely dashed] (uk.east) to[bend right=35] node[lab,right]{late report} (ec.east);
    \draw[arr,densely dashed] (uk.north)--(f.south);
    \draw[arr] (v.north) to[bend right=12] node[lab,right=1pt]{\textsc{Delay}} (dl.east);
    \draw[arr] (dl.west) to[bend right=12] node[lab,left=1pt]{requeue} (p.north);
    \draw[arr] (v.south) to[bend left=12] node[lab,right=1pt]{\textsc{Degrade}} (dg.east);
    \draw[arr] (dg.west) to[bend left=12] node[lab,left=1pt]{requeue} (p.south);
    \begin{scope}[on background layer]
      \node[draw=xorange!60,dashed,fill=xorange!3,rounded corners=4pt,inner sep=8pt,fit=(c)(p)(v)(a)(dl)(dg)] (live) {};
      \node[draw=xpurple!70,densely dotted,thick,rounded corners=3pt,inner sep=4pt,fit=(v)(dl)(dg)] (timed) {};
    \end{scope}
    \draw[arr,xred,thick] ($(live.north east)+(-0.9,0)$) |- node[lab,pos=0.7,above=2pt,text=xred]{\textsc{Revoke}} (r.west);
    \draw[arr,xred,thick] (timed.south) -- node[lab,right=1pt,pos=0.45,text=xred]{deadline exceeded} (ex.north);
    % legend
    \begin{scope}[shift={(0,-5.5)},font=\scriptsize]
      \node[st=xorange,minimum width=0.5cm,minimum height=0.28cm,text height=0pt,text depth=0pt,inner sep=0pt,label={[font=\scriptsize,text=xink]right:XAIR: validation, authorization}] at (0,0) {};
      \node[st=xblue,minimum width=0.5cm,minimum height=0.28cm,text height=0pt,text depth=0pt,inner sep=0pt,label={[font=\scriptsize,text=xink]right:released by the gateway}] at (7.4,0) {};
      \node[st=xgreen,minimum width=0.5cm,minimum height=0.28cm,text height=0pt,text depth=0pt,inner sep=0pt,label={[font=\scriptsize,text=xink]right:effect at the controller}] at (14.0,0) {};
      \node[st=xgold,minimum width=0.5cm,minimum height=0.28cm,text height=0pt,text depth=0pt,inner sep=0pt,label={[font=\scriptsize,text=xink]right:effect failed or unknown}] at (0,-0.75) {};
      \node[st=xred,minimum width=0.5cm,minimum height=0.28cm,text height=0pt,text depth=0pt,inner sep=0pt,label={[font=\scriptsize,text=xink]right:stopped before release}] at (7.4,-0.75) {};
      \node[st=xpurple,minimum width=0.5cm,minimum height=0.28cm,text height=0pt,text depth=0pt,inner sep=0pt,label={[font=\scriptsize,text=xink]right:deferred (delayed, degraded)}] at (14.0,-0.75) {};
      \node[term=xmuted,minimum width=0.5cm,minimum height=0.28cm,text height=0pt,text depth=0pt,inner sep=0pt,label={[font=\scriptsize,text=xink]right:double border: terminal state}] at (0,-1.5) {};
      \draw[draw=xorange!60,dashed,fill=xorange!3,rounded corners=2pt] (7.15,-1.65) rectangle (7.65,-1.35); \node[anchor=west] at (7.75,-1.5) {before commit or release};
      \draw[draw=xpurple!70,densely dotted,thick,rounded corners=2pt] (13.75,-1.65) rectangle (14.25,-1.35); \node[anchor=west] at (14.35,-1.5) {timed: the deadline can expire it};
    \end{scope}
  \end{tikzpicture}%
  }
  \caption{Intent lifecycle. Authorization, commit, release and effect are distinct states: an authorized intent is released after the gate (optimistic mode) or after its commit (atomic mode), and only an actuator or controller report confirms the effect, which stays unknown until a late report (dashed) resolves it. A refused gate, commit or release guard withholds the intent, and delayed and degraded intents are requeued as pending. Thick arrows mark the nominal path; colors and borders are explained in the legend.}
  \label{fig:fsm}
\end{figure}
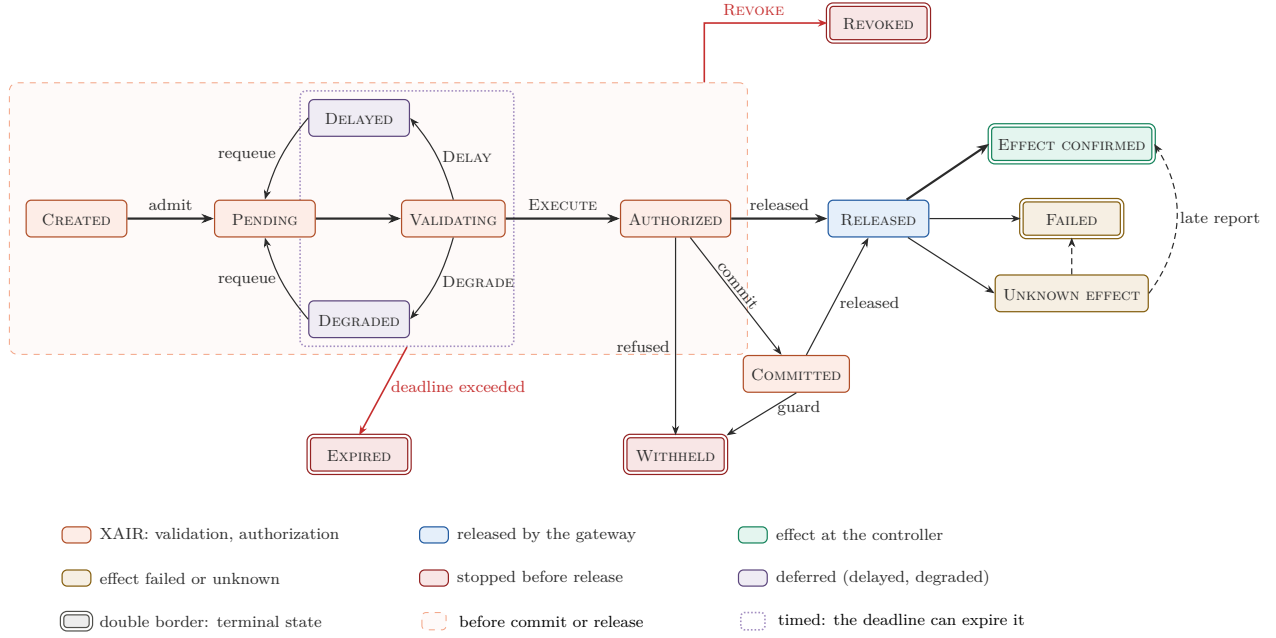

In XAIR, the instants defined in \Cref{sec:execution-gap} are mapped onto two components, namely XAIR itself and the actuator gateway. At $t_v$, XAIR validates the intent against a single snapshot and records both the global version of that snapshot and the read-set version $\nu_R(t_v)$ of the intent. At $t_g$, the gateway reads the snapshot again and rechecks temporal validity, $P$, $S$ and the version, and at $t_m$ it passes the command to the middleware. For each intent, XAIR keeps separate records of the policy outcome (\textsc{Execute}, \textsc{Delay}, \textsc{Degrade} or \textsc{Revoke}), the lifecycle state, a machine-readable reason, the release decision and the reported effect (\Cref{fig:fsm}). We model authorization, commit, release and effect as distinct states. A successful validation leads to \textsc{Authorized} and an atomic commit to \textsc{Committed}. When the gateway reports the middleware call, the intent moves to \textsc{Released}, or to \textsc{Withheld} if the gate, the commit or a release guard refused it. Only a report from the actuator or the controller can move it to \textsc{Effect confirmed} or \textsc{Failed}, and an intent that was released without such feedback stays in \textsc{Unknown effect}. Being authorized or committed is thus never taken as evidence of execution. The distinction matters in practice, because an audit that mixed the two states would not be able to tell a withheld command from a stale release, and for the same reason SER and FPR are always computed from the release decision and never from a validation label.

What happens at $t_v$ depends on the outcome of validation. An intent that is too old, or whose constraints or preconditions do not hold, is revoked (\textsc{Revoke}); one whose deadline has passed while its freshness window has not ends up as \textsc{Expired}. Among valid intents, those aimed at a busy target are delayed (\textsc{Delay}), those that declare a degradation policy are degraded and get a transformed payload (\textsc{Degrade}), and all the others are executed (\textsc{Execute}). A producer that resubmits the same \texttt{RESUME}, for example after a timeout, simply gets the retained record back, and nothing is released twice; conflicting intents on the same target are resolved according to their priority. Both \textsc{Delay} and \textsc{Degrade} send the intent back for another validation pass under the same identifier (a degraded intent is revalidated immediately, and its transformed payload is returned for release). These two outcomes are part of the contract, but we do not evaluate them in this paper. A supervisory revoke remains effective until the intent is committed or released, since the atomic commit refuses a revoked intent and the optimistic gate, which reads the intent's state together with its snapshot, withholds it. In the optimistic mode, however, a revoke that arrives after the gate's read, i.e., in $\ioc{t_g}{t_m}$, can no longer stop the release; in that case the release reported afterwards is rejected and recorded in the audit trail.

We provide two release modes because a deployment has to choose between a cheaper recheck and a stronger audit trail. In the \emph{optimistic} mode, the gateway rechecks the intent at $t_g$ and then releases it, which leaves the interval $\ioc{t_g}{t_m}$ unprotected. In the \emph{atomic} mode, the gateway asks XAIR to commit the authorization instead. XAIR compares the version, checks the elapsed time against ${d-\varepsilon}$ on the store clock and appends an authorization record to an actuation log, in a single store operation serialized with every context update. The gateway releases the intent only once this authorization has been committed. In this mode a release guard can additionally recheck the deadline right before the middleware call, logging any release it withholds as a tombstone after the committed record, and the commit can reserve a time margin for $\ioc{t_c}{t_m}$.

Alternatively, an actuator-side consumer reads the log in commit order from a durable offset and applies each record through an idempotent actuator. Assume durable writes on a single store primary and an actuator that persists its deduplication record together with the effect. A gateway crash after $t_c$ then loses no command, and a retried commit or a replayed record causes no second effect, as in log-based event processing~\cite{kleppmann2019online}. A record applied late may no longer be admissible, so the consumer can also recheck its version and deadline at apply time and withhold it (off by default).

In either mode the gateway can release through a version-conditional command at a controller that owns part of the context. It passes the controller's version of that state, as observed at validation (atomic mode) or at the gate read (optimistic mode), together with the absolute deadline ${t_d+d-\varepsilon}$. The controller executes the command only if both still hold in the scan that executes it (the PLC program of the evaluation checks the version only). Alternatively, a controller-local interlock (\Cref{sec:execution-gap}) checks the predicates on the controller's own state in that scan.

\subsection{Architecture and Prototype}
\label{sec:architecture}
\label{sec:implementation}

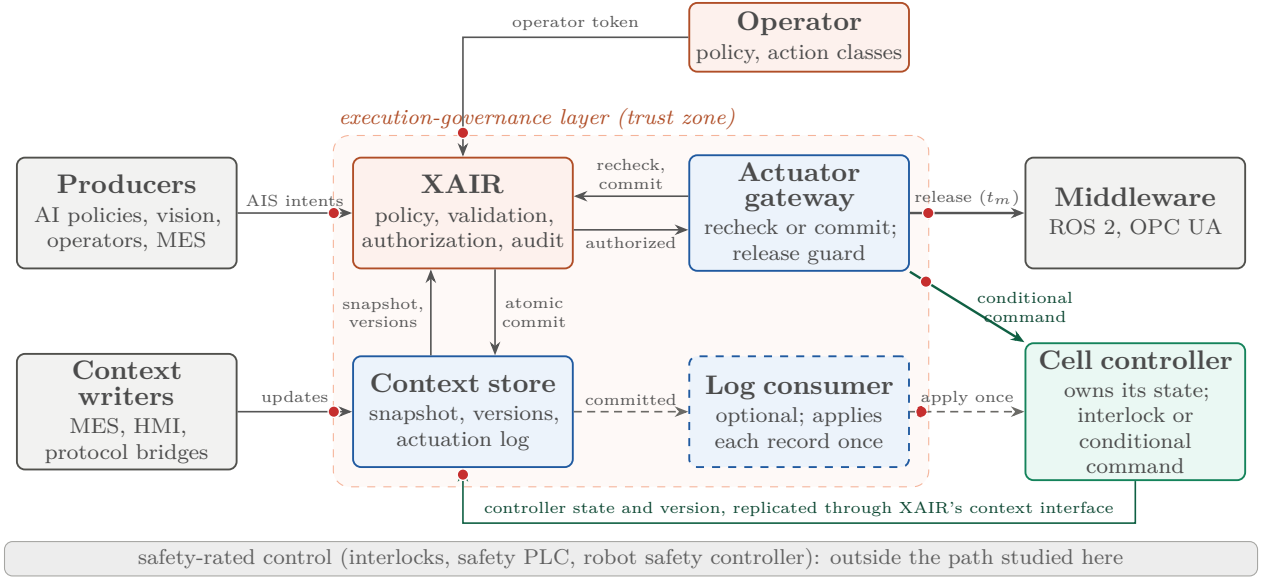
\begin{figure}[!t]
\centering
\resizebox{\linewidth}{!}{%
\begin{tikzpicture}[x=1cm,y=1cm,font=\scriptsize,
  comp/.style={draw=#1!75!black,fill=#1!9,rounded corners=3pt,line width=0.7pt,text width=2.7cm,minimum height=1.45cm,align=center,inner sep=2.5pt,text=xink!85,font=\scriptsize,execute at begin node={\hyphenpenalty=10000\exhyphenpenalty=10000}},
  comp/.default=xmuted,
  arr/.style={-{Stealth[length=1.8mm,width=1.4mm]},line width=0.6pt,xink!80},
  darr/.style={arr,densely dashed,xmuted},
  lab/.style={font=\tiny,inner sep=1.3pt,align=center,text=xink!80},
  ttl/.style={font=\footnotesize\bfseries,text=xink},
  cdt/.style={circle,fill=xred,draw=white,line width=0.5pt,inner sep=1.5pt}]
  \node[comp] (prod) at (1.6,0) {{\footnotesize\bfseries Producers}\\[1pt] AI policies, vision, operators, MES};
  \node[comp=xorange] (xair) at (6.0,0) {{\footnotesize\bfseries XAIR}\\[1pt] policy, validation,\\ authorization, audit};
  \node[comp=xblue] (gw) at (10.4,0) {{\footnotesize\bfseries Actuator gateway}\\[1pt] recheck or commit;\\ release guard};
  \node[comp] (mw) at (14.8,0) {{\footnotesize\bfseries Middleware}\\[1pt] ROS~2, OPC~UA};
  \node[comp] (wr) at (1.6,-2.6) {{\footnotesize\bfseries Context writers}\\[1pt] MES, HMI,\\ protocol bridges};
  \node[comp=xblue] (store) at (6.0,-2.6) {{\footnotesize\bfseries Context store}\\[1pt] snapshot, versions, actuation log};
  \node[comp=xblue,dashed] (cons) at (10.4,-2.6) {{\footnotesize\bfseries Log consumer}\\[1pt] optional; applies\\ each record once};
  \node[comp=xorange,minimum height=0.9cm] (op) at (10.4,2.3) {{\footnotesize\bfseries Operator}\\[1pt] policy, action classes};
  \node[comp=xgreen] (ctl) at (14.8,-2.6) {{\footnotesize\bfseries Cell controller}\\[1pt] owns its state;\\ interlock or\\ conditional command};
  \begin{scope}[on background layer]
    \node[fill=xorange!4,draw=xorange!50,dashed,rounded corners=6pt,inner sep=7pt,fit=(xair)(gw)(store)(cons)] (gov) {};
  \end{scope}
  \node[font=\scriptsize\itshape,text=xorange!75!black,anchor=south west,inner sep=2pt] at (gov.north west) {execution-governance layer (trust zone)};
  % producers and writers
  \draw[arr] (prod)--node[lab,above=1pt]{AIS intents}node[cdt,pos=0.84]{}(xair);
  \draw[arr] (wr)--node[lab,above=1pt]{updates}node[cdt,pos=0.84]{}(store);
  \draw[arr] (op.west)-|node[lab,pos=0.25,above=1pt]{operator token}node[cdt,pos=0.9]{}(xair.north);
  % XAIR <-> store
  \draw[arr] ($(store.north)+(-0.42,0)$)--node[lab,left=1pt]{snapshot,\\versions}($(xair.south)+(-0.42,0)$);
  \draw[arr] ($(xair.south)+(0.42,0)$)--node[lab,right=1pt]{atomic\\commit}($(store.north)+(0.42,0)$);
  % gateway <-> XAIR
  \draw[arr] ($(gw.west)+(0,0.22)$)--node[lab,above=1pt]{recheck,\\ commit}($(xair.east)+(0,0.22)$);
  \draw[arr] ($(xair.east)+(0,-0.22)$)--node[lab,below=1pt]{authorized}($(gw.west)+(0,-0.22)$);
  % release paths
  \draw[arr,line width=0.9pt] (gw)--node[lab,above=1pt]{release ($t_m$)}node[cdt,pos=0.16]{}(mw);
  \draw[arr,xgreen!55!black,line width=0.9pt] (gw.south east)--node[lab,right=2pt,pos=0.5,text=xgreen!40!black]{conditional\\ command}node[cdt,pos=0.14]{}(ctl.north west);
  % atomic mode: log consumer
  \draw[darr] (store)--node[lab,above=1pt]{committed}(cons);
  \draw[darr] (cons)--node[lab,above=1pt]{apply once}node[cdt,pos=0.1]{}(ctl);
  % controller state replicated back to the store
  \draw[arr,xgreen!55!black] (ctl.south)--++(0,-0.55)-|node[lab,pos=0.25,above=1pt,text=xgreen!40!black]{controller state and version, replicated through XAIR's context interface}node[cdt,pos=0.94]{}(store.south);
  % safety layer
  \draw[xmuted!60,fill=xmuted!12,rounded corners=3pt] (0.0,-4.75) rectangle (16.4,-4.3);
  \node[font=\scriptsize,text=xink!75] at (8.2,-4.525) {safety-rated control (interlocks, safety PLC, robot safety controller): outside the path studied here};
\end{tikzpicture}}
\caption{XAIR data path. Producers submit AIS intents, and XAIR validates them against the versioned snapshot. The gateway rechecks or commits through XAIR before releasing, or releases a version-conditional command that the controller owning the context executes only if its version is unchanged. Context writers, including the controller's replicated state, update the store through XAIR. The dashed box is the trust zone, and red dots mark the conduits that the deployment must authenticate, since the prototype protects only the policy interface.}
\label{fig:architecture}
\end{figure}

\Cref{fig:architecture} shows the architecture. XAIR runs as an edge service that validates incoming intents against the JSON Schema, performs temporal checks on the server side using the UTC clock with a declared clock bound, and maintains a versioned snapshot of the context. Only the actuator gateway (or, in the atomic mode, the consumer of the actuation log) can turn an authorization into a command. Because the commit, the release report and the effect report each advance the lifecycle record, near misses, i.e., cases in which validation succeeds but the recheck blocks the release, remain observable. Replayed reports are handled idempotently, and conflicting reports are rejected. We keep the lifecycle records in XAIR's memory, whereas the idempotency table, the context, the versions, the committed authorizations and the actuation log are held by the store (restarts are discussed in \Cref{sec:discussion}). If the store cannot be reached, XAIR marks the context as untrusted and rejects all intents, so that an outage fails closed. A failed audit after release, on the other hand, cannot undo the call to the gateway, which is why downstream safety logic remains authoritative. Finally, the optimistic mode should be understood as an application-level optimistic concurrency check~\cite{kung1981optimistic} rather than a two-phase commit. It narrows the TOCTOU race~\cite{bishop1996toctou} but does not eliminate it.

Our prototype uses Redis as the context store. The context, its global version and a map of per-path versions are kept in a single document, which is updated through an optimistic compare-and-set transaction; in this way a reader always obtains values and versions that were written together. XAIR derives the read-set of an intent from its parsed predicates and returns $\nu_R(t_v)$ along with its decision. For the atomic commit we rely on a server-side script, because Redis runs such a script without interleaving any other command; the script checks the version and the elapsed time against the store clock, refuses duplicates, and appends the authorization record to the log. Predicate-truth versions are a little more expensive to maintain. Whenever an update touches a path, the registered predicates that read that path are re-evaluated in the same transaction, and a predicate's version moves only if its truth value flips. We retire predicates that have not been used for a retention period, so that this cost does not grow without bound. In addition, every access to the store returns time bounds in the monotonic clock shared by the processes running on the same kernel, which allows the gateway to place $t_g$, $t_c$ and any context write relative to its own $t_m$.

For the evaluation we built three additional components. The first is a set of adapter-local caches that mirror the snapshot with different coherence policies, namely refresh before every validation, push invalidation, intentional staleness, and an authoritative read followed by a version recheck. Since they use the same temporal validator and predicate evaluator as XAIR, they only differ in which snapshot they read and in the recheck they perform, and none of them applies the server-side policy. The second is an external baseline based on a policy engine, Open Policy Agent (OPA)~1.21~\cite{opa2025}. OPA evaluates the same predicates, translated into a generic rule, against a replica of XAIR's context, which is written after every update together with its versions and its trust flag. The gateway queries OPA when an intent arrives and again just before release, either with the predicates only or with the read-set condition of \Cref{eq:validity-p}; we use this second configuration to test whether versioning is a property of the check rather than of XAIR. The third component is an emulated cell controller, implemented as an OPC~UA server that owns the line state and its version and executes queued commands with a 10\,ms scan. It offers an unconditional command, a controller-local interlock and a version-conditional command, and it replicates every state change to XAIR, as an OPC~UA subscription would. On the PLC runtime, the same version check is implemented as a Structured Text program running on OpenPLC~v3 and accessed over Modbus~TCP.

The guarantees of XAIR are conditional, and their conditions are spread over the model, the prototype and the deployment. \Cref{tab:guarantees} brings them together, with the failure that follows when one is violated. XAIR cannot check most of them at run time. A writer that changes the line state without going through the context interface, for example, is simply invisible to every gate, so the table is also a list of what a deployment has to provide before the results of \Cref{sec:evaluation} apply to it.

\begin{table}[!htbp]
\caption{Guarantees of AIS/XAIR, the assumptions each of them needs, and what happens if an assumption is violated.}
\label{tab:guarantees}
\centering
\gridtable\scriptsize
\gridzebra{2}
\begin{tabularx}{\linewidth}{|L{3.1cm}|Y|Y|}
\hline
\gridhead\textbf{Guarantee} & \textbf{Requires} & \textbf{If violated} \\
\hline
Gate or commit blocks every change ordered before it & every transition is written to the store through XAIR's context interface, with versions advancing atomically & a change never written is invisible, and the intent is released stale \\
\hline
Deadline respected at the last check ($t_g$ or $t_c$) & clock offset within $\varepsilon$ & a command older than $d$ passes the check \\
\hline
Release within the deadline ($t_m$) & the above; in the atomic mode a release guard or a commit margin; in the optimistic mode nothing closes $\ioc{t_g}{t_m}$ & a command is released older than $d$, by at most that interval \\
\hline
Committed authorizations are exactly ordered & one store primary serializes commits and updates & with failover or unfenced replicas a commit can be ordered against stale state \\
\hline
Truth versions block an undone change (A--B--A) & every intermediate state is written; registration and snapshot read are one step; half the predicate retention exceeds the longest freshness window or deadline & an unwritten flip goes unseen; a retired predicate fails closed \\
\hline
No stale effect on controller-owned state & (A1)--(A5), with versions never reused & an old witness can match a reused version; our PLC program meets (A4) only within one run \\
\hline
No lost or repeated command in the actuation log & durable writes on one primary; the actuator persists its deduplication record with the effect & a crash can lose a command or repeat its effect \\
\hline
A resubmitted intent is released at most once & XAIR records every admitted identifier for the retention window & the store loses the table or the window passes, and the identifier can be released again \\
\hline
Producers cannot drop a mandatory predicate & only the operator can change the policy & a client that writes the policy can remove any predicate \\
\hline
XAIR never revokes a stop & the policy assigns the action type to the stop class & an unclassified stop is gated like any other intent \\
\hline
A stop blocks the earlier intents of its target & the stop time is written before the stop is released, by no other writer; clocks within $\varepsilon$; one XAIR process & an intent past its last check is released after the stop; a failed write leaves earlier intents unfenced \\
\hline
\end{tabularx}
\end{table}

\section{Evaluation}
\label{sec:evaluation}

The evaluation asks where the last check must run, and what each placement costs, through five research questions. \textbf{RQ1}: what do contextual preconditions change, compared with a guard that has none? \textbf{RQ2}: where must context be read once a remote writer can change it? \textbf{RQ3}: what does each release path leave unprotected, and which component must perform the last check? \textbf{RQ4}: what does the version condition cost when the context is updated often, and how often does drift occur in a production log? \textbf{RQ5}: what does the runtime cost on edge processors, and how does it scale and fail?

\begin{figure}[!t]
  \centering
  \includegraphics[width=\linewidth]{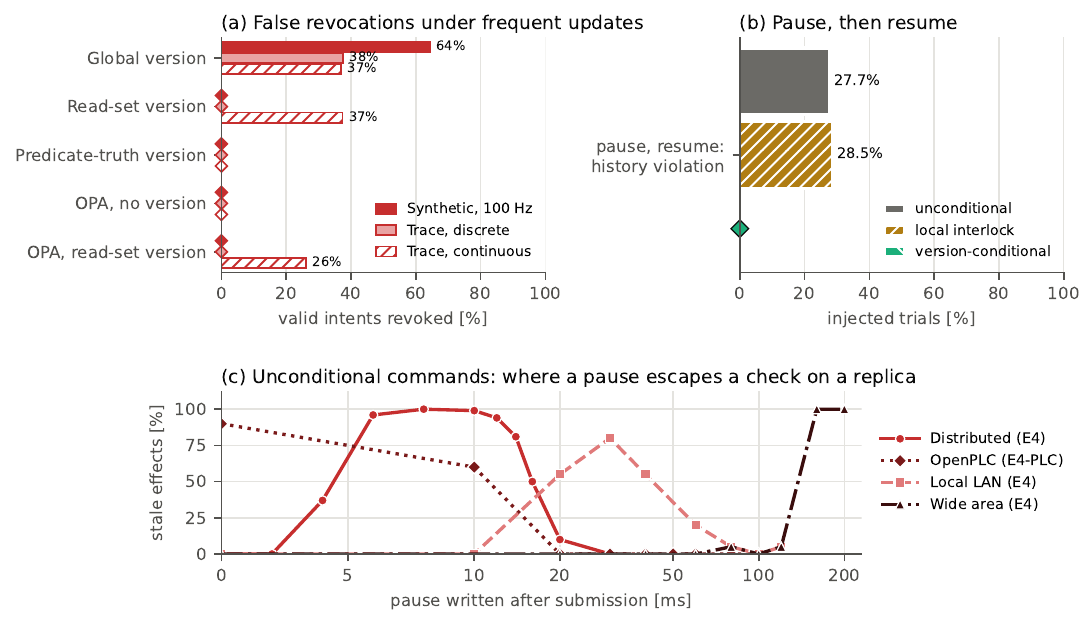}
  \caption{Key results. (a) Valid intents revoked by each gate under updates at 100\,Hz (\expChurn) and under a recorded hydraulic trace (\expTrace); diamonds mark zero. (b) History violations at the controller after a pause that was undone, by command type (\expCtrlAba). (c) Stale effects of unconditional commands by the offset of the pause after submission, on a symmetric-log axis: the unprotected interval that a check on a replica leaves on each testbed (\expCtrl, \expPlc).}
  \label{fig:keyresults}
\end{figure}

The results that decide the design are collected in \Cref{fig:keyresults}, after the setup; the subsections take the questions in turn.

\subsection{Setup}

We designed the evaluation as a controlled ablation. The same AIS payloads and drift schedules are replayed under a series of guards that all share XAIR's temporal validator and predicate evaluator, so that the only difference between them is where, and how, the context is checked. Direct release and freshness-only validation serve as failure floors. The guards we compare are local caches with different coherence policies, a read-through check, XAIR itself, and a policy engine with and without the read-set condition (\Cref{sec:architecture}); at the controller, a local interlock acts as the baseline for the version-conditional command. In a drift trial the preconditions of the intent are invalidated after the intent has been built, while in a valid trial they stay true; every guard is therefore judged on both SER and FPR, and we observe both at the gateway boundary, since an intent that was authorized at $t_v$ may still be withheld at $t_g$. 

All experiments declare ${\varepsilon=0}$, that is, they take the clocks of the producer and of the checker to agree. A few terms are used in one sense only. A window is always the freshness window $w$, whereas the unprotected interval is the time after a check in which a change goes unseen, a replica is any copy of the context that can lag its owner, such as the store snapshot, a cache or the engine's copy, and the gate is the component that makes the last check before release, by recheck at $t_g$ or by commit at $t_c$. An intent is obsolete when its assumption no longer holds, a trial has drift when we inject such a change, and only a release or an effect is called stale.

Two kinds of results follow. In conformance tests the outcome is fixed by the design: a gate that compares versions cannot release an intent whose version has moved, and a command evaluated in the scan of the controller cannot act on a state it has not seen, so these experiments check only that the version and the deadline arrive intact and that valid commands still execute, and their zeros carry no discovery. In measurements the outcome could have been different: the unprotected interval that a check on a replica leaves, the false revocations of each version scope, the accuracy of \Cref{eq:fpr}, and the cost on edge hardware. For the experiments that depend on timing, the replication unit is a campaign: we ran five independent campaigns and report their ranges, without drawing any population-level inference. The experiments are listed in \Cref{tab:suites}, and their protocols, per-condition results and per-trial data are released with the artifact, and the file experiments.yaml maps our identifiers to the names in its scripts and data.

\begin{figure}[!t]
  \centering
  \resizebox{\linewidth}{!}{%
  \begin{tikzpicture}[x=1cm,y=1cm,
    chip/.style={draw=xink!70,fill=white,rounded corners=2pt,minimum width=1.5cm,minimum height=0.5cm,font=\scriptsize,align=center,inner sep=1pt},
    role/.style={draw=#1!75!black,fill=#1!12},
    mach/.style={draw=#1!70!black,fill=#1!9,rounded corners=4pt,inner sep=5pt,minimum height=2.5cm},
    mach/.default=xblue,
    one/.style={minimum width=3.9cm},
    half/.style={minimum width=1.85cm},
    cont/.style={draw=xink!60,densely dashed,rounded corners=3pt,inner sep=2.5pt},
    net/.style={draw=xred,thick,-{Stealth[length=1.5mm]},densely dashed},
    ptitle/.style={font=\footnotesize\bfseries,anchor=north west,inner sep=0pt},
    pnote/.style={font=\tiny\itshape,text=xink,align=center}]
    % every panel is 4.3 cm wide; its machine boxes are 2.5 cm high
    % ---------------- row 1
    \begin{scope}[shift={(0,0)}]  % S
      \node[ptitle] at (0,0) {S \normalfont single host};
      \node[chip,role=xmuted] (Sp) at (1.15,-0.85) {producers}; \node[chip,role=xblue] (Ss) at (2.75,-0.85) {store};
      \node[chip,role=xorange] (Sx) at (1.15,-1.55) {XAIR};      \node[chip,role=xblue] (Sg) at (2.75,-1.55) {gateway};
      \node[chip,role=xgreen] (Sc) at (1.15,-2.25) {controller};
      \begin{scope}[on background layer]\node[mach=xmuted,one] at (1.95,-1.55) {};\end{scope}
      \node[pnote,anchor=west] at (2.0,-2.25) {all on loopback};
    \end{scope}
    \begin{scope}[shift={(4.6,0)}]  % R
      \node[ptitle] at (0,0) {R \normalfont reserved cores};
      \node[chip,very thick,role=xmuted] (Rp) at (1.15,-0.85) {producers}; \node[chip,very thick,role=xblue] (Rs) at (2.75,-0.85) {store};
      \node[chip,very thick,role=xorange] (Rx) at (1.15,-1.55) {XAIR};      \node[chip,very thick,role=xblue] (Rg) at (2.75,-1.55) {gateway};
      \begin{scope}[on background layer]\node[mach=xmuted,one] at (1.95,-1.55) {};\end{scope}
      \node[pnote] at (1.95,-2.25) {own CPU set each, no controller};
    \end{scope}
    \begin{scope}[shift={(9.2,0)}]  % D
      \node[ptitle] at (0,0) {D \normalfont distributed};
      \node[chip,role=xmuted] (Dp) at (1.15,-0.85) {producers}; \node[chip,role=xblue] (Ds) at (2.75,-0.85) {store};
      \node[chip,role=xorange] (Dx) at (1.15,-1.55) {XAIR};      \node[chip,role=xblue] (Dg) at (2.75,-1.55) {gateway};
      \node[chip,role=xgreen] (Dc) at (1.15,-2.25) {controller}; \node[chip,role=xmuted] (Do) at (2.75,-2.25) {OPA};
      \begin{scope}[on background layer]
        \node[mach=xmuted,one] at (1.95,-1.55) {};
        \foreach \c in {Dp,Ds,Dx,Dg,Dc,Do} {\node[cont,fit=(\c)] {};}
      \end{scope}
      \node[pnote,anchor=north] at (1.95,-2.85) {one container each, 0.5\,ms per interface};
    \end{scope}
    \begin{scope}[shift={(13.8,0)}]  % PLC
      \node[ptitle] at (0,0) {PLC \normalfont OpenPLC};
      \node[chip,role=xmuted] (Op) at (1.15,-0.85) {producers}; \node[chip,role=xblue] (Os) at (2.75,-0.85) {store};
      \node[chip,role=xorange] (Ox) at (1.15,-1.55) {XAIR};      \node[chip,role=xblue] (Og) at (2.75,-1.55) {gateway};
      \node[chip,role=xgreen] (Oc) at (2.75,-2.25) {OpenPLC};
      \draw[net] (Og.south) -- (Oc.north);
      \begin{scope}[on background layer]
        \node[mach=xmuted,one] at (1.95,-1.55) {};
        \node[cont,fit=(Oc)] {};
      \end{scope}
    \end{scope}
    % ---------------- row 2
    \begin{scope}[shift={(0,-3.7)}]  % L
      \node[ptitle] at (0,0) {L \normalfont local LAN};
      \node[chip,minimum width=1.4cm,role=xmuted] (Lp) at (0.9,-0.85) {producers}; \node[chip,minimum width=1.4cm,role=xblue] (Ls) at (0.9,-1.55) {store}; \node[chip,minimum width=1.4cm,role=xorange] (Lx) at (0.9,-2.25) {XAIR};
      \node[chip,minimum width=1.4cm,role=xblue] (Lg) at (3.4,-1.2) {gateway};    \node[chip,minimum width=1.4cm,role=xgreen] (Lc) at (3.4,-1.9) {controller};
      \draw[net] (1.85,-1.55) -- (2.45,-1.55);
      \node[font=\tiny,inner sep=1pt,align=center] at (2.15,-1.2) {LAN\\6.6\,ms};
      \begin{scope}[on background layer]\node[mach=xmuted,half] at (0.9,-1.55) {}; \node[mach=xmuted,half] at (3.4,-1.55) {};\end{scope}
      \node[pnote,anchor=north] at (0.9,-2.85) {Raspberry Pi~4}; \node[pnote,anchor=north] at (3.4,-2.85) {Apple M4};
    \end{scope}
    \begin{scope}[shift={(4.6,-3.7)}]  % P
      \node[ptitle] at (0,0) {P \normalfont wide-area VPN};
      \node[chip,minimum width=1.4cm,role=xmuted] (Pp) at (0.9,-0.85) {producers}; \node[chip,minimum width=1.4cm,role=xblue] (Ps) at (0.9,-1.55) {store}; \node[chip,minimum width=1.4cm,role=xorange] (Px) at (0.9,-2.25) {XAIR};
      \node[chip,minimum width=1.4cm,role=xblue] (Pg) at (3.4,-1.2) {gateway};    \node[chip,minimum width=1.4cm,role=xgreen] (Pc) at (3.4,-1.9) {controller};
      \draw[net] (1.85,-1.55) -- (2.45,-1.55);
      \node[font=\tiny,inner sep=1pt,align=center] at (2.15,-1.2) {VPN\\62\,ms};
      \begin{scope}[on background layer]\node[mach=xmuted,half] at (0.9,-1.55) {}; \node[mach=xmuted,half] at (3.4,-1.55) {};\end{scope}
      \node[pnote,anchor=north] at (0.9,-2.85) {server}; \node[pnote,anchor=north] at (3.4,-2.85) {Apple M4};
    \end{scope}
    \begin{scope}[shift={(9.2,-3.7)}]  % B
      \node[ptitle] at (0,0) {B \normalfont embedded board};
      \node[chip,role=xmuted] (Bp) at (1.15,-1.05) {producers\\[-1pt]\tiny core 0}; \node[chip,role=xblue] (Bg) at (2.75,-1.05) {gateway\\[-1pt]\tiny core 1};
      \node[chip,role=xorange] (Bx) at (1.15,-2.0) {XAIR\\[-1pt]\tiny core 2};       \node[chip,role=xblue] (Bs) at (2.75,-2.0) {store\\[-1pt]\tiny core 3};
      \begin{scope}[on background layer]\node[mach=xmuted,one] at (1.95,-1.55) {};\end{scope}
      \node[pnote,anchor=north] at (1.95,-2.85) {Pi~4 or VisionFive~2, no controller};
    \end{scope}
    \begin{scope}[shift={(13.8,-3.7)},font=\scriptsize]  % legend
      \node[ptitle] at (0,0) {legend};
      \node[chip,role=xmuted,minimum width=0.45cm,minimum height=0.22cm] at (0.22,-0.62) {}; \node[anchor=west] at (0.6,-0.62) {producers};
      \node[chip,role=xblue,minimum width=0.45cm,minimum height=0.22cm] at (0.22,-1.0) {}; \node[anchor=west] at (0.6,-1.0) {store, gateway};
      \node[chip,role=xorange,minimum width=0.45cm,minimum height=0.22cm] at (0.22,-1.38) {}; \node[anchor=west] at (0.6,-1.38) {XAIR};
      \node[chip,role=xgreen,minimum width=0.45cm,minimum height=0.22cm] at (0.22,-1.76) {}; \node[anchor=west] at (0.6,-1.76) {controller};
      \draw[cont] (0,-2.26) rectangle (0.45,-2.04); \node[anchor=west] at (0.6,-2.15) {container};
      \draw[net] (0,-2.55) -- (0.45,-2.55); \node[anchor=west] at (0.6,-2.55) {network link};
    \end{scope}
  \end{tikzpicture}}
  \caption{Where the components ran in each testbed configuration: the producers with the load generator, the context store, XAIR, the gateway and the cell controller. Chip colors follow the other figures (XAIR orange, store and gateway blue, controller green, producers grey), grey boxes are machines and dashed borders are containers. The controller is an emulated OPC UA controller, except in PLC, and R and B run no controller. Only D includes a policy-engine container (OPA).}
  \label{fig:testbeds}
\end{figure}
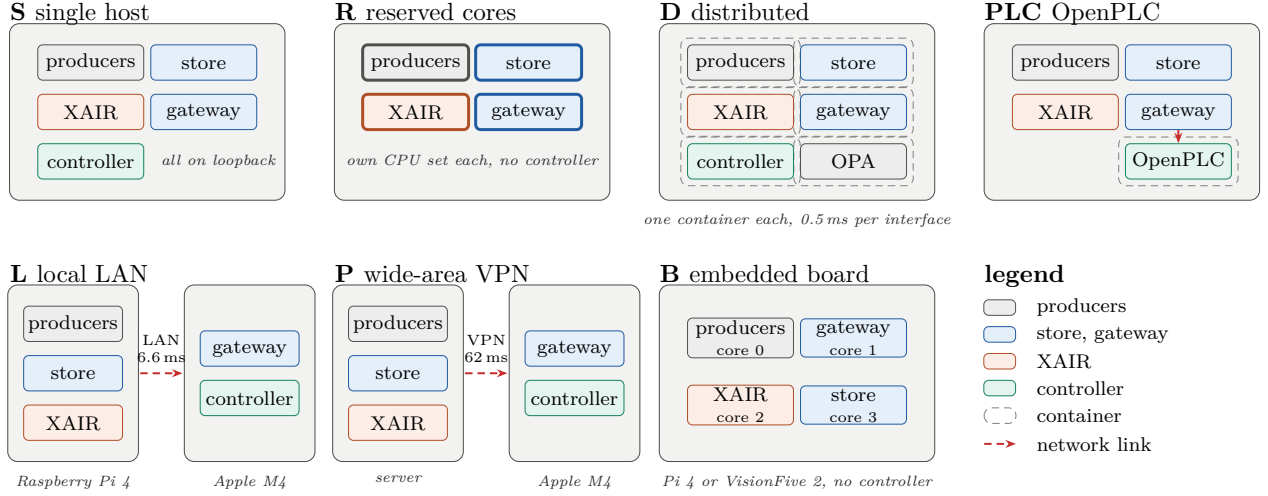

Six testbed configurations, S, D, R, P, L, and B, span the placements a deployment may choose, from loopback to a wide-area link and a single embedded board. \Cref{fig:testbeds} shows where each component ran, and \Cref{tab:testbeds} gives the hardware and the constraints of each configuration. All run Python~3.12 with FastAPI and Redis~7.4. The latencies of the single host (S) are orders of magnitude only, since every component shares one machine. The distributed testbed (D) uses one shared monotonic clock and hosts five campaigns and three sensitivity variants (a jitter of $2\pm1$\,ms, shared CPUs, and a phase-locked trace replay), whereas on P and L, without a shared clock, only store versions order events. The two embedded boards (B) count as one configuration, and the OpenPLC runtime (PLC) is not part of it: an OpenPLC~v3 container on the server executes the controller program as Structured Text in a 10\,ms task over Modbus~TCP.

\begin{table}[!tp]
\caption{Hardware and constraints of the testbed configurations. The server is a dual-socket AMD EPYC~7313 machine (32 cores, 64 threads, 512\,GB), and the Apple machine has an M4 chip. The Pi has a Cortex-A72 CPU and a real-time kernel, the VisionFive~2 a SiFive U74 (RISC-V) CPU. TCP connect times are medians.}
\label{tab:testbeds}
\centering
\gridtable
\renewcommand{\arraystretch}{1.0}\setlength{\tabcolsep}{3pt}
\gridzebra{2}
\begin{tabularx}{\linewidth}{|L{1.25cm}|L{3.1cm}|Y|L{3.0cm}|}
\hline
\gridhead\textbf{Config.} & \textbf{Machines} & \textbf{Placement and isolation} & \textbf{Network} \\
\hline
S & server & all processes on one host, no pinning & loopback \\
\hline
R & server & XAIR, gateway, producers and a Redis container on disjoint CPU sets & loopback \\
\hline
D & server & one Docker container per component (Redis~7, OPA~1.21, XAIR, gateway, controller, producers) on a bridge, each with its own CPU set & $0.5\pm0.1$\,ms of shaped delay per interface, TCP connect 1.2\,ms \\
\hline
P & server and Apple M4 machine & producers, store and XAIR on the server; gateway and controller on the Apple machine & wide-area VPN, TCP connect 62\,ms \\
\hline
L & Raspberry Pi~4 and Apple M4 machine & producers, store and XAIR on the Pi; gateway and controller on the Apple machine & local LAN over Wi-Fi, TCP connect 6.6\,ms \\
\hline
B & Raspberry Pi~4 or VisionFive~2, four cores each & four components on one board, one core each (producers, gateway, XAIR, store); no controller & loopback \\
\hline
PLC & server, OpenPLC~v3 container & as S, with the controller replaced by an OpenPLC task of 10\,ms & Modbus TCP \\
\hline
\end{tabularx}
\end{table}

\begin{table}[!tp]
\caption{Experiments by research question, with the testbed codes of \Cref{fig:testbeds,tab:testbeds}; ``offline'' marks an analysis of the log. Variants are labels after the number. \expPause, \expStore (with \expStoreAba), \expCtrl (with \expCtrlAba) and \expChurn were also run on P and L, whereas \expPlc, \expCtrlChurn, \expStoreStop and \expChurnScale ran only on PLC and S, and \expStoreDeadline and \expTrace only on D. Campaigns are independent repetitions of a suite (D: five, boards count as one), and experiments.yaml in the artifact maps every identifier to its scripts and data files.}
\label{tab:suites}
\centering
\gridtable
\renewcommand{\arraystretch}{1.0}\setlength{\tabcolsep}{3pt}
\gridzebra{2}
\begin{tabularx}{\linewidth}{|L{2.1cm}|c|L{1.6cm}|L{2.0cm}|Y|Y|}
\hline
\gridhead\textbf{Experiment} & \textbf{RQ} & \textbf{Testbeds} & \textbf{Campaigns} & \textbf{Protocol} & \textbf{Finding} \\
\hline
\expPause & 1 & S D P L & 1 & pause before release; 100 drifted and 100 valid per guard; \expClasses: three further action classes; \expRos: simulated ROS~2 cell; \expPolicy: precondition omitted, with or without policy & floors release every obsolete intent, contextual guards none; an omitted precondition is released unless the operator mandates it \\
\hline
\expCache & 2 & S D & 1 & remote write 0--500\,ms before submission & a stale cache releases every obsolete intent; XAIR none \\
\hline
\expStore & 3 & S D P L B & D 5, others 1 & write 0--100\,ms after validation; 30 per offset; \expStoreDeadline: $d=100$\,ms, gate delay 75--105\,ms; \expStoreAba: pause and resume while the gate waits (D P L); \expStoreStop: a stop of the target while the check waits (S) & writes before the check blocked, 1--4\,ms remain; an atomic commit needs a guard or a margin; only versioned gates block an undone change; a stop fences earlier intents \\
\hline
\expCtrl & 3, 4 & S D P L PLC & D 5, others 1 & pause 0--200\,ms after submission, resumed or not (\expCtrlAba); \expPlc: OpenPLC runtime; \expCtrlChurn: three controller variables at 0--100\,Hz each, 800 valid commands per cell (S) & checks in the scan stop stale effects; only a version also after a resume; a version for all controller state leaves 37--100\% of valid commands unexecuted, one for the line state none \\
\hline
\expChurn & 4 & S D P L B & D 5, others 1 & writers at 0--500\,Hz; \expTrace: hydraulic trace; \expChurnScale: $10$--$10^4$ registered predicates, updates at 20 and 100\,Hz (S) & global versions revoke valid intents; truth versions none, but the store lag grows with the registry \\
\hline
\expLog & 4 & offline, D & 1 & job-shop log; paired replay through five gates & drift rare within a minute; no obsolete decision released by XAIR \\
\hline
\expEdge & 5 & B & 1 & 3$\times$3000 sequential intents; updates at 0--100\,Hz, and 3$\times$400 trials per cell at 20--100\,Hz & one pinned core validates 148--168 intents/s \\
\hline
\expScale & 5 & R S D & 1 & $10^4$ intents; 1--50 producers; mixed drift and valid intents; up to $10^4$ predicates & all released up to 50 producers; truth-version cost grows with touched paths \\
\hline
\expFaults & 5 & S D & 1 & \expNetwork: network impairment; \expCrash: crashes of the committer, the consumer and the store & no stale release; no lost or duplicated command \\
\hline
\end{tabularx}
\end{table}

\subsection{Where context is read (RQ1, RQ2)}

Preconditions help only when a guard evaluates them, and when the operator can make the important ones mandatory. With a pause injected before release, direct release and freshness-only validation released every obsolete \texttt{RESUME}, and none of the contextual guards released one or revoked a valid intent, on any testbed where they ran (\expPause). The same holds for three further action classes and in a simulated ROS~2 cell, where an independent subscriber on the actuator topic agreed with the gateway on every trial (\expClasses, \expRos). A producer that omits the line-state precondition is released after every pause, because XAIR checks what is declared; an operator policy that makes the precondition mandatory blocks all such releases (\expPolicy).

Once a remote writer can change the context, where it is read decides the outcome (\expCache). A cache that is never refreshed released every obsolete intent, a cache refreshed by push notification only those submitted before its notification arrived, and read-through and XAIR none.

\subsection{What a check on a replica leaves open (RQ3)}

A recheck in the store blocks every change ordered before it, but not one that arrives after it (\expStore). Invalidating writes injected after validation were ordered against the check by the store version they created, and every write ordered before the check was blocked, in both release modes. What remains is the interval between the check and the middleware call, 1--2\,ms at the median on the distributed testbed and about 4\,ms on the embedded boards. The atomic commit makes the authorization exact without shortening this interval. On the distributed testbed it lengthened the median validation-to-release time of valid intents by 0.01--0.19\,ms per campaign, paired by campaign (4.34--4.49 against 4.26--4.33\,ms), and it holds the deadline at release only with a release guard or a commit margin (\expStoreDeadline). The injected writes did not land in so short an interval, so the experiment measures its width; the controller experiments show that changes inside such an interval do reach the actuator.

Only a versioned check detects a change that was undone before it (\expStoreAba). When the line is paused and resumed while the gateway waits, every predicate holds again at the check (\Cref{tab:e10aba}). Gates without versions released all such intents, on the distributed testbed and on both physical nodes, whereas read-set and predicate-truth versions blocked all of them, in XAIR and in the policy engine once it was given the versions. The protection therefore comes from the version and moves with it: given the same versions, the policy engine behaved like XAIR. The blocking is implied by the design, and all 125 valid control trials per gate still passed.

\begin{table}[tbp]
\caption{\expStoreAba: history-violating intents released after a predicate change was undone before the check, by gate (distributed: 250 per gate over five campaigns; physical: 50 per gate and node whose flip falls between validation and the check in store order; the policy engine runs on the distributed testbed only).}
\label{tab:e10aba}
\centering
\gridtable
\gridzebra{3}
\begin{tabular}{|l|c|c|c|c|}
\hline
\gridhead& \multicolumn{2}{c|}{\textbf{Distributed}} & \multicolumn{2}{c|}{\textbf{Physical, released}} \\
\hhline{|~|-|-|-|-|}
\gridhead\textbf{Gate} & \textbf{Released} & \textbf{Controls} & \textbf{Wide area} & \textbf{Local LAN} \\
\hline
XAIR, read-set version & 0/250 & 125/125 & 0/50 & 0/50 \\ \genrule
XAIR, predicate-truth version & 0/250 & 125/125 & 0/50 & 0/50 \\ \genrule
XAIR, predicate-only & 250/250 & 125/125 & 50/50 & 50/50 \\ \genrule
XAIR atomic, read-set version & 0/250 & 125/125 & 0/50 & 0/50 \\ \genrule
XAIR atomic, predicate-truth version & 0/250 & 125/125 & 0/50 & 0/50 \\ \genrule
Policy engine (OPA), predicates only & 250/250 & 125/125 & -- & -- \\ \genrule
Policy engine (OPA), read-set version & 0/250 & 125/125 & -- & -- \\ \genrule
 
\end{tabular}
\end{table}

A stop is the case in which an intent already in flight must not be released, and in XAIR a stop is never rechecked, takes no lock and enters no actuation log (\expStoreStop). XAIR therefore fences the intents of the target with the time of each stop, as described with the action classes. We submitted a \texttt{RESUME} with its gate or commit delayed by 40\,ms and a stop of the same target 0--60\,ms later, 330 injected trials per release mode, and counted the \texttt{RESUME} as stale when it was released after the stop. Without the fence (the ablation condition), 238 of 330 trials in the optimistic mode and 245 of 330 in the atomic mode were stale, in every trial that the stop reached while the check was waiting. With the fence, none of 330 was stale in either mode (95\% upper bound 1.2\%), and every one of the 100 valid control trials per cell was still released. The \texttt{RESUME} that were released with the fence went out before the stop was released. A stop that arrives after the release is not stopped by this, and the fence covers only the intents on one target of one XAIR process.

At the controller, however, checking XAIR's replica of the state turns out not to be enough (\expCtrl). In this experiment an emulated OPC~UA controller owns the line state, runs a scan every 10\,ms and replicates its state to XAIR asynchronously, and the operator pause is written at the controller shortly after the intent has been submitted (\Cref{tab:e21}). Because XAIR's replica lags behind the controller, unconditional commands acted on a paused line in about half of the injected trials. This rate mostly depends on where the pauses fall (\Cref{fig:keyresults}c). A pause that reaches XAIR before its check is blocked, and one that arrives after the effect does no harm, but a pause that falls in between usually escapes, in up to 100\% of the trials on the distributed testbed and in 80\% on the LAN node. On the distributed testbed this interval lies a few milliseconds after submission: at least half of the pauses written 6--16\,ms after submission escaped. On the LAN node the same share escaped for pauses 20--40\,ms after submission, and on the wide-area node from 160\,ms on, up to the largest offset we tried (200\,ms). With a controller-local interlock or a version-conditional command, on the other hand, no stale effect remained (\Cref{fig:keyresults}b; 0 of 1100 injected trials per release path in \Cref{tab:e21}), and every valid command was still executed. With 0 of 1100 trials the 95\% upper bound is 0.35\% (Wilson). The measured results are the unprotected interval that a check on a replica leaves and the history violations of the interlock in the pause--resume columns.

The two differ when the pause is undone before the effect (\expCtrlAba). With a resume shortly after each pause, the interlock found a running line and executed 28.5\% of these commands after the interruption in the optimistic mode (285 of 1000) and 29.7\% in the atomic mode, about 29\% over both, against 27.7\% for the unconditional command in the optimistic mode. The version-conditional command refused all of them, because the version had advanced (\Cref{fig:keyresults}b). On the physical nodes the pattern for pauses repeats, while pause--resume pairs rarely fit between check and effect.

\begin{table}[tbp]
\caption{\expCtrl: effects on a paused, controller-owned line, by release path: stale effects after a pause, and stale effects and history violations after a pause and resume; distributed testbed (five campaigns) and the two physical nodes. Pause--resume pairs rarely fit between the check and the effect on the wide-area node, which has no such column. Every valid control trial executed.}
\label{tab:e21}
\centering
\gridtable
\gridzebra{4}
\begin{tabular}{|l|c|c|c|c|c|c|}
\hline
\gridhead& \multicolumn{3}{c|}{\textbf{Distributed}} & \textbf{Wide area} & \multicolumn{2}{c|}{\textbf{Local LAN}} \\
\hhline{|~|-|-|-|-|-|-|}
\gridhead& pause & \multicolumn{2}{c|}{pause--resume} & pause & pause & pause--resume \\
\hhline{|~|-|-|-|~|-|-|}
\gridhead\textbf{Release path} & \textbf{Stale} & \textbf{Stale} & \textbf{History} & \textbf{Stale} & \textbf{Stale} & \textbf{History} \\
\hline
optimistic, unconditional & 567/1100 & 277/1000 & 277/1000 & 42/180 & 44/180 & 4/140 \\ \genrule
atomic, unconditional & 569/1100 & 286/1000 & 281/1000 & 37/180 & 51/180 & 5/140 \\ \genrule
optimistic, local interlock & 0/1100 & 0/1000 & 285/1000 & 0/180 & 0/180 & 1/140 \\ \genrule
atomic, local interlock & 0/1100 & 0/1000 & 297/1000 & 0/180 & 0/180 & 3/140 \\ \genrule
optimistic, version-conditional & 0/1100 & 0/1000 & 0/1000 & 0/180 & 0/180 & 0/140 \\ \genrule
atomic, version-conditional & 0/1100 & 0/1000 & 0/1000 & 0/180 & 0/180 & 0/140 \\ \genrule
 
\end{tabular}
\end{table}

The OpenPLC experiment (\expPlc) was meant to rule out an artifact of our emulated controller. There, the controller logic is implemented as a Structured Text program executed in a 10\,ms task, and the results are essentially the same (\Cref{fig:keyresults}c): about one in six unconditional commands acted on a paused line, and no command did so with the interlock or with the version-conditional command (0 of 180 per release path in a single campaign, with a 95\% upper bound of 2.1\%). The program checks only the version, since the deadline is enforced at the gateway, the runtime is containerized, and the ``effect'' is a register write rather than a physical output. This experiment therefore confirms the scan boundary of a software PLC, not the behavior of a physical controller.

\subsection{What versioning costs (RQ4)}

Only a global version turns unrelated updates into revocations (\expChurn). With background writers changing an unrelated field or rewriting the line state with its current value, every revocation is a false positive (\Cref{fig:keyresults}a). A global version revoked 12\% of valid intents at 20\,Hz and 64\% at 100\,Hz on the distributed testbed, and the rate grows with the gate latency on every testbed. Read-set and predicate-truth versions, and the policy engine with or without versions, revoked none.

With continuous signals, however, read-set versions are not sufficient (\expTrace). To show this, we replayed a public recording of a hydraulic test rig~\cite{helwig2015hydraulic,uci2018hydraulic} as notifications from an OPC~UA subscription. The recording never touches the line state, but the pressure changes value in almost every notification (\Cref{fig:keyresults}a). As a result, an intent requiring the pressure to stay below a threshold, which it never actually crosses, was revoked in 37\% of the cases at a 10\,ms publishing interval, i.e., as often as with a global version. Predicate-truth versions, by contrast, revoked no intent at all, since the truth value of the predicate never changed. The policy engine with its read-set rule revoked 26\% of the continuous intents and none of the discrete ones. These runs are small in one respect: every intent registered two or three predicates, and the store applied about 77 updates per second at the 10\,ms interval. How the store behaves with a large registry is a separate question, which the scaling runs below address.

The same question arises at the controller (\expCtrlChurn). The version-conditional command compares a version that the controller owns, and if that version advances on every change of controller state, it behaves like a global version in the scan. We tested this with a controller that owned three fast variables besides the line state, each changing at 20, 50 or 100\,Hz (60, 150 and 300 changes per second in total), while the valid commands read only the line state. With one version for all controller state, the controller refused 30\%, 89\% and all of the commands that the gateway released, and the gateway itself had already revoked 10\%, 25\% and 46\% of all valid commands because the mirrored version moved, so 37\%, 92\% and 100\% of them were not executed. With a version that advances only on the line state, none of 800 valid commands was refused at any rate (95\% upper bound 0.5\%). The protection against an undone pause did not depend on the scope: neither version let a history violation through in 200 pause--resume trials per cell.

The run is on the single host, so absolute times are only indicative. A scan took 1.6--5.4\,\textmu s on average and at most 15.1\,\textmu s at the 99th percentile in every cell, against a period of 10\,ms. \Cref{eq:fpr}, using the median end-to-end time of each cell (7--9\,ms) as a proxy for $\Delta_g$, which is not logged in this experiment, predicts 40\%, 73\% and 88\% for the three rates, to be compared with the 37\%, 92\% and 100\% not executed. It follows the trend but underestimates the higher rates, where XAIR's replica of the version lags more: the delay that reproduces the controller's refusals grows from 6 to at least 15\,ms. A controller with fast state therefore has to scope its version to what the command reads. The emulated controller does this at the price of one comparison in the scan, and we have not repeated it on a PLC runtime.

How often drift occurs depends on when decisions are taken and on what counts as invalidation (\expLog, \Cref{tab:e20}). In the event log of a job shop~\cite{levy2014production}, drift within a minute is rare, below 1\% of decisions, unless decisions are taken the moment a machine becomes idle. For decisions that wait an hour, it affects 5--43\% of decisions, depending on how the plant schedules, and a change of product alone up to a tenth. A paired replay of a month of the log on the distributed testbed checks the gates on this drift: each of \TwentyPDecThirty{} decisions per latency went through all five gates, and \TwentyPObsThirty{} at a latency of 30 plant minutes and \TwentyPObsTwoForty{} at 240 were obsolete for every gate. Direct release released all of them and every XAIR gate none, which bounds their stale-release rate below 9.9\% and 2.5\% (Wilson 95\%). A global version also revoked 41/\TwentyPValThirty{} and 39/\TwentyPValTwoForty{} valid decisions, read-set and truth versions none, which bounds their false-revocation rate below 0.21\% (Wilson 95\%). The two latencies see the same update stream (2.81\,Hz), so the global version costs about the same at both. The valid decisions are fewer at 240 minutes (1852 against 1963) simply because more decisions are obsolete by then. The replay compresses plant time 4320-fold and covers one log, so we read it as a sanity check of the controlled results against production drift and would not transfer its rates to other plants.

\begin{table}[tbp]
\caption{\expLog: drift in a production event log. Top: decisions [\%] whose machine left the idle state, or whose next step was for another part or a breakdown, within the decision-to-release latency (offline). Bottom: paired live replay, obsolete decisions released / obsolete and valid decisions revoked / valid, per gate and latency; the \TwentyPDiscThirty{} and \TwentyPDiscTwoForty{} decisions whose ground truth differed between gates are excluded.}
\label{tab:e20}
\centering
\gridtable
\gridzebra{2}
\begin{tabular}{|l|c|c|c|c|}
\hline
\gridhead\textbf{Decision model; invalidation} & 1\,min & 5\,min & 30\,min & 1\,h \\
\hline
Decisions uniform over idle time & 0.13 & 0.58 & 3.0 & 5.4 \\ \genrule
\quad idle intervals $\le$\,8\,h only & 0.97 & 3.9 & 19 & 32 \\ \genrule
One decision when a machine becomes idle & 14 & 21 & 35 & 43 \\ \genrule
Uniform; next step is a different part & 0.046 & 0.21 & 1.1 & 2.2 \\ \genrule
Uniform; next step is a breakdown & 0.0022 & 0.0073 & 0.035 & 0.059 \\ \genrule
 
\end{tabular}
\vspace{2.5ex}
\gridzebra{3}
\begin{tabular}{|l|c|c|c|c|}
\hline
\gridhead& \multicolumn{2}{c|}{\textbf{30 plant minutes}} & \multicolumn{2}{c|}{\textbf{240 plant minutes}} \\
\hhline{|~|-|-|-|-|}
\gridhead\textbf{Gate} & stale & false rev. & stale & false rev. \\
\hline
Direct & 35/35 & 0/1963 & 147/147 & 0/1852 \\ \genrule
XAIR, global & 0/35 & 41/1963 & 0/147 & 39/1852 \\ \genrule
XAIR, read-set & 0/35 & 0/1963 & 0/147 & 0/1852 \\ \genrule
XAIR, predicate-truth & 0/35 & 0/1963 & 0/147 & 0/1852 \\ \genrule
XAIR atomic, predicate-truth & 0/35 & 0/1963 & 0/147 & 0/1852 \\ \genrule
 
\end{tabular}
\end{table}

\subsection{Cost on edge processors, scaling, and faults (RQ5)}

Since the design places XAIR on an edge node, we also measured it on two embedded boards (\expEdge, \Cref{tab:e25}). We pinned the XAIR process to one core of the four-core board and gave the store, the gateway and the load generator one core each. In this setting XAIR validated 148--168 sequential intents/s on the Raspberry Pi~4 and on the VisionFive~2, which is about a sixth of what the same single-process runtime achieves on reserved server cores. Each request (a validation without release) took about 6\,ms end to end at the median, and XAIR used less than 100\,MB of resident memory. Because the gate is slower on these boards, a global version also becomes more expensive, and at 100\,Hz it revoked almost every valid intent, whereas read-set and truth versions again revoked none.

\begin{table}[tbp]
\caption{\expEdge: XAIR on embedded boards (Raspberry Pi~4, Cortex-A72, real-time kernel; VisionFive~2, SiFive U74, RISC-V), each process pinned to one core: sequential throughput, median end-to-end (E2E) latency, valid intents revoked by a global version at 20 and 100\,Hz (100 trials per cell here; the repeat with 1200 is in \Cref{sec:discussion}), median validation-to-gate interval $\Delta_g$ at 100\,Hz, and resident memory (RSS) of XAIR. Read-set and truth versions revoked none.}
\label{tab:e25}
\centering
\gridtable
\begin{tabular}{|l|c|c|c|c|c|}
\hline
\gridhead& \textbf{Intents/s} & \textbf{E2E [ms]} & \textbf{Global FPR, 20 / 100\,Hz [\%]} & \textbf{$\Delta_g$ [ms]} & \textbf{RSS [MB]} \\
\hline
Raspberry Pi 4 & 167--168 & 5.7 & 14 / 99 & 8.6 & 96 \\ \genrule
VisionFive 2 & 148--152 & 6.4 & 26 / 95 & 9.9 & 93 \\ \genrule
 
\end{tabular}
\end{table}

Outside the edge boards, we also checked how the runtime scales (\expScale) and how it behaves under faults (\expNetwork, \expCrash). On reserved server cores it validates 909--956 intents/s, and closed loops with up to 50 producers released every intent. When drifted and valid intents were mixed over three seeds, all decisions were correct. A read-through cache and the policy engine are faster; the policy engine, in particular, keeps no lifecycle or release records and does not read the store. The cost of predicate-truth versions is highest for updates that touch many registered predicates. With a thousand predicates, one per context path, the store applied 78--136 updates/s when each update changed at most a tenth of those paths, and 44--64 updates/s when it changed all of them (three repetitions on a shared host).

These rates are of the same order as the 77 updates/s of the trace, and a truth flip protects an intent only once its update has been applied, so we measured the lag end to end through XAIR and Redis (\expChurnScale). With 10, 1000 and 10000 registered predicates, and updates at 20 and 100\,Hz that each changed a tenth of the paths, the document layout of the registry kept up with 100\,Hz only at 10 predicates. It applied 53 updates/s at 1000 predicates and 4.9 updates/s at 10000, with a median update latency of 2.3\,ms at 10 predicates, 16.5\,ms at 1000 and 206\,ms at 10000. The hash layout sustained 100\,Hz at 1000 predicates (7.1\,ms) and 10.8 updates/s at 10000 (93\,ms). None of the 120 valid control trials per cell was revoked, and none of the 546 flips that fell between validation and the check was released. The lag, however, moved the flips out of that interval: of 100 per cell, 87 and 91 (document and hash, 1000 predicates, 100\,Hz) and all 100 at 10000 predicates no longer fitted between validation and the check, so at that scale the protocol cannot test undone changes. We did not measure the stale-release rate inside the lag itself, which lengthens the interval in which an undone change goes unseen. Under network impairment no drifted intent was released, but a freshness window shorter than the transport delay also led to the revocation of valid intents.

Finally, the actuation log lost no command and produced no second effect under emulated and real crashes of the committer, the consumer and the store, although this alone does not prove exactly-once physical effects. The lifecycle, batch-conflict and malformed-input regression suites also all passed; these suites ship with the artifact and are not counted as experiments.

\FloatBarrier
\section{Discussion}
\label{sec:discussion}
\label{sec:implications}

Our results suggest that the risk depends on \emph{which} state a guard observes, much more than on how fast the guard is. In our experiments the fastest contextual guard, i.e., the cache that is never refreshed, failed on every drift trial, while XAIR, which is the slowest, failed on none. A latency figure alone therefore says little about semantic safety unless it comes with the consistency class, the observation boundary and the drift protocol under which it was measured. For multi-producer cells this means that an authoritative read, a central gate, or a hybrid design is needed; in the latter, local checks can take part of the load while XAIR maintains the shared version and the audit trail. We condense these results into a design guide in \Cref{tab:placement}.

\begin{table}[tbp]
\caption{Design guide: which changes to the context the last check blocks, by where it runs and which version it compares (\Cref{sec:execution-gap,sec:evaluation}). A change can be committed before the check, after it, or committed and undone before the check (A--B--A). The last column names the updates that revoke valid intents; their rate times the gate latency sets the cost (\Cref{eq:fpr}).}
\label{tab:placement}
\centering
\gridtable
\begin{tabularx}{\linewidth}{|l|l|c|c|c|Y|}
\hline
\gridhead & & \multicolumn{3}{c|}{\textbf{Change to the context}} & \\
\hhline{|~|~|-|-|-|~|}
\gridhead\multirow{-2}{*}{\textbf{Last check}} & \multirow{-2}{*}{\textbf{Version}} & \textbf{before check} & \textbf{after check} & \textbf{undone} & \multirow{-2}{*}{\textbf{Revoked by}} \\
\hline
Local cache & none & no & no & no & -- \\
\hline
\multirow{4}{*}{Store gate} & none & blocked & no & no & -- \\
\cline{2-6}
 & global & blocked & no & blocked & every update \\
\cline{2-6}
 & read-set & blocked & no & blocked & read-set value changes \\
\cline{2-6}
 & predicate truth & blocked & no & blocked & predicate truth flips \\
\hline
\multirow{2}{*}{Policy engine (replica)} & none & blocked$^{d}$ & no & no & -- \\
\cline{2-6}
 & read-set, given by XAIR & blocked$^{d}$ & no & blocked & read-set value changes \\
\hline
Atomic commit & read-set or truth & blocked$^{a}$ & detected$^{b}$ & blocked & as its version \\
\hline
\multirow{2}{*}{Controller scan} & none (interlock) & blocked & blocked$^{c}$ & no & -- \\
\cline{2-6}
 & controller version & blocked & blocked$^{c}$ & blocked & changes of the state it covers \\
\hline
\end{tabularx}

\vspace{2pt}
\raggedright\scriptsize $^{a}$\,exactly ordered in the store; $^{b}$\,not blocked, but detected and audited from store versions; $^{c}$\,only for context the controller owns; $^{d}$\,only once the update has been replicated to the engine: XAIR commits the update to the store first and then copies it, counts a failed copy and lets the next update overwrite it, so a change can be ordered in the store and still be invisible to the engine.
\end{table}

The false revocations we measured on every testbed are predicted reasonably well by \Cref{eq:fpr}. Using the measured gate latencies, the better of its two forms is within 3.5--6.7 percentage points of the single-host, distributed, trace and physical-node cells, and within 13 points on the embedded boards. On the boards we repeated the update-rate cells with 1200 trials each (three seeds, 20, 50 and 100\,Hz), and, with the better of the two forms of the equation, ten of the twelve deviations are underestimates, between 3.5 and 12.6 percentage points; the other two are overestimates of about one point. The largest is on the VisionFive~2 at 20\,Hz with writes to an unrelated field. Averaging the prediction over the measured distribution of $\Delta_g$ instead of taking its median moved it by 2.4 points at most, and the distribution is narrow (95th percentile about 1.5 times the median at most). The gap therefore does not come from the median or from the tail. It probably comes from a part of the version-read-to-comparison interval that the validation-to-gate time does not cover, and we have not isolated it. We use the model as a first-order guide on the boards.

Since the cost is governed by $\lambda\Delta_g$, slower gates, such as those on embedded boards or across a wide-area link, lose more valid intents to a global version, so the slower the edge processor, the more the scope of the version matters. With read-set versions, $\lambda$ reduces to the rate of value changes on $R(I)$, which is negligible for supervisory state but not for a threshold on a fast signal. Predicate-truth versions reduce it further, to the rate of truth flips, and still detect A--B--A violations, at the price of re-evaluating the registered predicates on every update that touches them; a policy engine that was given the same versions behaved in the same way. A--B--A violations are not an academic corner case on the shop floor, as anyone who has seen a guard door opened and closed while a command waits will know. Whether the command should then proceed is a policy decision, and the version scope is where this decision is encoded.

As for the release paths, in the optimistic mode a change that falsifies $P$ or $S$ escapes the gate with probability of roughly $\lambda_R\Delta_r$. With the intervals we measured and one such change per second, this is of the order of ${10^{-3}}$ per release, to which the deadline adds the probability that the elapsed time crosses $d$ within $\Delta_r$. The atomic commit does not remove this term. What it does is make the authorization exact, order every later update with respect to it, and allow a consumer to withhold commands that were invalidated after their commit, while the interval itself remains, whatever the store does. It can only be closed by a check performed where the context is owned, in the same step that produces the effect, and so only for the part of the read-set that the executing controller owns. When the gate sees the controller's state only through replication, the interval is much longer, and what matters is where it starts, which depends on replication and transport (\Cref{sec:evaluation}). Choosing which check to run at the controller is, again, a choice of version scope, just as at the gate (\expCtrlChurn), and an interlock that cannot see an undone change leaves the history violations that only a version excludes. If the controller checks only the version, as our PLC program does, the deadline is enforced at the gateway rather than at the effect.

How often the gap matters in practice is a different question. In the production log we analyzed, drift within a minute (which is far longer than any software release path) is rare, unless decisions are taken at the very moment a machine becomes free. Decisions that have to wait, whether for a language model, for an operator's confirmation or in a scheduling queue, see their machine change state in a few percent to almost half of the cases within an hour. One log is of course not enough to fix the rate for another plant, but it does show that the rate depends on scheduling at least as much as on latency. Fast-changing context such as telemetry or controller-owned state can drift within milliseconds, and not all misses have the same cost, since a resume released onto a line that an operator has just paused may move an arm toward that operator. We therefore see the mechanism as aimed at actions that either have a high consequence or wait for a long time. For actions that do neither, a global version is probably not worth the false revocations it causes.

The version-conditional command maps naturally onto existing industrial standards. In OPC~UA it can be exposed as a method whose input arguments carry the expected version and the absolute deadline, which is what the controller of \expCtrl does. In IEC~61499 it would be a function block whose request event carries the expected version as data~\cite{vyatkin2011iec}, and in a PLCopen motion program the same comparison would guard the execute input of the motion block. When assets are described by an Asset Administration Shell~\cite{iec63278aas}, AIS paths can refer directly to its submodel properties. Finally, the trust model that we only partly address in this paper fits the zones and conduits of IEC~62443~\cite{iec62443zones}, with XAIR, the store and the gateway forming one zone, and with the conduits through which producers, context writers and the operator reach that zone (\Cref{fig:architecture}) required to authenticate them.

In our view, fail-closed revocation makes sense for actions that increase activity, for example resume, move or grasp, and this is why the action classes of \Cref{sec:action-abstraction} exempt supervisory stops from every check that could revoke them, while emergency stops stay on the safety-rated path. Which actions belong to which class is a decision for the operator and the safety engineer, and getting it wrong in either direction has a cost, since a stop that is not classified is gated like any other intent and a resume classified as a stop would bypass the gate. A second practical point concerns the size of the freshness windows, which has to take the transport delay into account. In \expNetwork, as soon as the delay exceeded the freshness window, the gate effectively became a denial-of-service mechanism, and for this reason $w$, $d$ and $\varepsilon$ should be treated as deployment parameters derived from the distribution of the delay between producer and gate.

\paragraph{Threats to validity}
\emph{Construct.} Deterministic experiments can show that an effect exists, but not how often it matters in a real plant. In particular, the absence of stale effects for controller-side checks is a conformance result, not a measurement (Section~5, Setup). The rates we measured on the unconditional path, for instance, depend on where the injected pauses happen to fall, and what can be transferred to other settings is therefore the extent of the unprotected interval rather than the pooled rate (the same applies to the store-check experiment, from which only the width of the interval should be carried over). We measured the cost of versions with synthetic updates and a single recorded test rig, and the prevalence of drift on a single production log under bounding decision models.
\emph{Internal.} Since injected writes need a network round trip to commit, none of them landed between the check and $t_m$ on the distributed testbed. On the physical nodes, the elapsed times computed at the gateway include the offset between two NTP-synchronized clocks, and the deadline guarantee with ${\varepsilon=0}$ assumes this offset to be small compared with freshness windows of one to five seconds. We did not record the offset during the campaigns and measured it afterwards against public servers: the producer host was within 4\,ms and the Apple machine within about 55\,ms (with an uncertainty of the same order from the round trip), and the Raspberry Pi~4 held 33\,\textmu s under its time daemon. An offset of that size changes the outcome of a decision only if its elapsed time at the gate lies within the offset of the bound, and the elapsed times we measure are tens of milliseconds against freshness windows of seconds. It would matter for freshness windows of the order of the transport delay, as in \expNetwork. Assumption (A4) requires that versions are never reused across restarts, whereas our PLC program restarts its version at zero and wraps a 16-bit register after about 327\,s. No campaign restarted a controller and the deadlines are far shorter, so the results do not depend on this, but a deployment would need a persistent epoch incremented at every restart and a counter that cannot wrap within the longest deadline. After a restart, XAIR loses its lifecycle records, so the gateway fails closed on intents it no longer knows but cannot record the withholding; the idempotency table lives in the store, which we checked with a unit test and a server restart, not in an experiment. The versioned rule of the policy engine relies on per-path versions replicated by XAIR, and the throughput we report for it does not include that cost. The artifact gives a SHA-256 manifest for every data file, which its verification script checks, and one snapshot of the code that reproduces every subset with its default options, but it does not tie each campaign to the commit that produced it. Finally, the production log does not resolve latencies below one minute, and its live replay, paired across gates, contains 35 and 147 obsolete decisions per latency. This is enough to bound the stale-release rate of each XAIR gate, but not to rank the XAIR gates against each other.
\emph{Statistical.} The Wilson intervals we give for $0/n$ and $n/n$ are descriptive within a cell. For a zero count, the tables always give the denominator, and the 95\% upper bound is about $3.84/(n+3.84)$, e.g., 2.1\% for the 180 OpenPLC trials of each release path. Because trials within a campaign share timing and infrastructure, we treat campaign ranges as the measure of replication. Only the distributed testbed has five campaigns, while the physical, LAN, embedded, OpenPLC and controller update-rate (\expCtrlChurn) results each come from a single campaign.
\emph{External.} The distributed testbed emulates its nodes on a single host, and the physical nodes are consumer hardware connected over a wide-area VPN and a local LAN rather than over a wired plant network or a fieldbus. The controllers of \expCtrl and \expPlc are software controllers without physical inputs or outputs, so our controller results describe their scan boundary, and a field trial with a physical PLC and wired I/O on a plant network is still missing. In addition, there is a single XAIR process without redundancy or authentication. All components run on the same board in the embedded experiments, the actuators of \expCrash are stubs, the predicate-truth runs with a realistic update stream registered only a few predicates, the stale-release rate inside the lag of the store with many of them was not measured, only the lag, and the predicate grammar only supports conjunctions.

Taken together, these limitations mean that our conclusions should be read narrowly: what XAIR governs is semantic admissibility at release time, while physical safety, in the sense of torque limits, collision envelopes and certified stops, stays with the PLC or the robot safety controller, which XAIR does not try to replace. We also do not see AIS/XAIR as a substitute for formal runtime enforcement, and it does not subsume typed-plan firewalls or contract-grounded robot runtimes either~\cite{petel2026typedplan,salfity2026contractbt}; it addresses a different question, and in a real cell we would expect it to be combined with them.

\section{Conclusion}
\label{sec:conclusion}

In this paper we started from a simple observation, namely that a command can be correct when it is decided and wrong when it takes effect. We treated this execution gap as a question of systems architecture: we separated the instants between decision and effect, built the resulting model into AIS and XAIR, and measured, under declared ground truth, where in a cell the last check has to run.

Our evaluation supports a single design principle, according to which the last check belongs in the serialization domain that owns the context it reads, and the version it compares has to be scoped to what the action depends on. A check on a store replica blocked every change ordered before it, with or without an atomic commit, but it always left an interval before the effect. For context owned by the controller, this interval was closed only by a check in the scan of our software controllers, and only the version-conditional command also refused commands after a pause that had been undone, whatever the scope of its version. The number of valid intents revoked by unrelated updates depended on the scope of the version, regardless of which runtime compared it, and at the controller as well as at the gate. Finally, the runtime fits on a single core of an edge board, and we expect the mechanism to matter most for actions that wait for inference or confirmation, or that act on fast-changing, controller-owned state.

Before deployment, the version and deadline arguments of version-conditional commands should be standardized, for instance in OPC~UA companion specifications, the action classes of the policy should be derived from the asset model and reviewed in the safety case, lifecycle records should persist across restarts, producers and context writers should be authenticated, and field trials on a wired plant network, with a physical controller and real I/O, are the obvious next experiment.

\section*{CRediT authorship contribution statement}
\textbf{Luca D'Agati:} Conceptualization, Methodology, Software, Investigation, Formal analysis, Data curation, Visualization, Writing -- original draft.
\textbf{Giuseppe Tricomi:} Methodology, Software, Validation, Writing -- review \& editing.
\textbf{Fabio Orazio Mirto:} Methodology, Software, Validation, Writing -- review \& editing.
\textbf{Giovanni Merlino:} Conceptualization, Supervision.

\section*{Declaration of competing interest}
The authors declare that they have no known competing financial interests or personal relationships that could have appeared to influence the work reported in this paper.

\section*{Data availability}
The implementation, the experiment scripts and container definitions, the per-trial data behind every table and figure, the aggregation scripts, and a detailed-results document with protocols and per-condition tables are publicly available at \url{https://github.com/ProSoDiAC/LD1-execution-gap-xair}. The experiment identifiers of the paper are the ones that the reproduction script accepts, and the names of the files in the data are listed in experiments.yaml. The two public datasets used are available from their original repositories~\cite{uci2018hydraulic,levy2014production}.

\section*{Funding}
This research did not receive any specific grant from funding agencies in the public, commercial, or not-for-profit sectors.

\section*{Declaration of generative AI and AI-assisted technologies in the manuscript preparation process}
During the preparation of this work the authors used Claude (Anthropic) in order to review the manuscript and to edit its language and style. After using this tool, the authors reviewed and edited the content as needed and take full responsibility for the content of the published article.

\bibliographystyle{elsarticle-num}
\bibliography{references}

\clearpage
\section*{About the authors}

\newcommand{\authorbio}[3]{%
  \noindent\begin{minipage}[t]{0.19\textwidth}\vspace{0pt}%
    \includegraphics[width=\linewidth,height=3.0cm,keepaspectratio]{authors/#1}%
  \end{minipage}\hfill
  \begin{minipage}[t]{0.78\textwidth}\vspace{0pt}\small\raggedright
    \textbf{#2} #3
  \end{minipage}\par\bigskip}

\authorbio{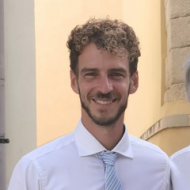}{Luca D'Agati}{is an assistant professor at the Department of Engineering, University of Messina, Italy, and a member of several research groups: the \textit{Mobile and Distributed System Lab} (MDSLAB), the \textit{Protocolized Software-defined and Decentralized intelligent Agentic Cyber-Physical Systems} (ProSoDiAC), and \textit{the National Interuniversity Consortium for Informatics} (CINI). His research concerns the Internet of Things (IoT), distributed systems and the Cloud-to-Edge continuum, in particular the secure and adaptive coordination of workloads across edge-to-cloud infrastructures, with applications in smart cities, industrial automation and telemedicine. He has taken part in national and EU projects.}

\authorbio{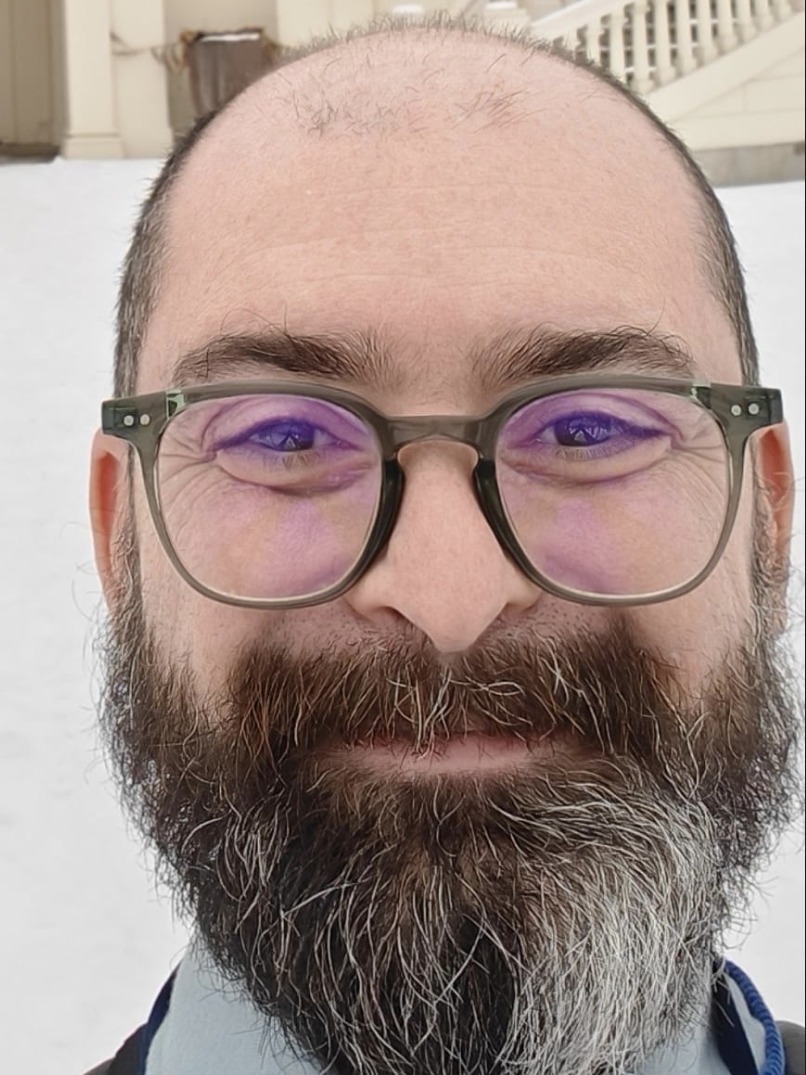}{Giuseppe Tricomi}{is an assistant professor at the Department of Engineering, University of Messina, Italy, and a member of several research groups: the \textit{Mobile and Distributed System Lab} (MDSLAB), the \textit{Protocolized Software-defined and Decentralized intelligent Agentic Cyber-Physical Systems} (ProSoDiAC), and \textit{the National Interuniversity Consortium for Informatics} (CINI). He received his international Ph.D. from the University of Messina in 2021 and was a researcher at the Institute of High-Performance Computing and Networking of the National Research Council of Italy until February 2025. His interests include cyber-physical systems, smart environments and the Cloud-to-Edge continuum. He has co-authored 47 papers, has taken part in national and EU projects, and belongs to the team leading the design of Stack4Things, an OpenStack-based Sensing-and-Actuation-as-a-Service framework.}

\authorbio{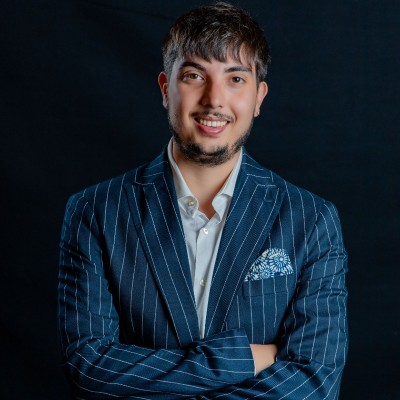}{Fabio Orazio Mirto}{is a Ph.D. student in Bioengineering Applied to Medical Sciences at the Department of Engineering, University of Messina, Italy, and a member of several research groups: the \textit{Mobile and Distributed System Lab} (MDSLAB) and the \textit{Protocolized Software-defined and Decentralized intelligent Agentic Cyber-Physical Systems} (ProSoDiAC). He received his Master's degree in Computer Engineering from Politecnico di Torino, Italy, in 2023. His research interests include cybersecurity, trustworthy and agentic AI, cyber-physical systems, edge-cloud computing, and the secure orchestration of distributed infrastructures, with emphasis on healthcare, IoT and safety-critical systems.}

\authorbio{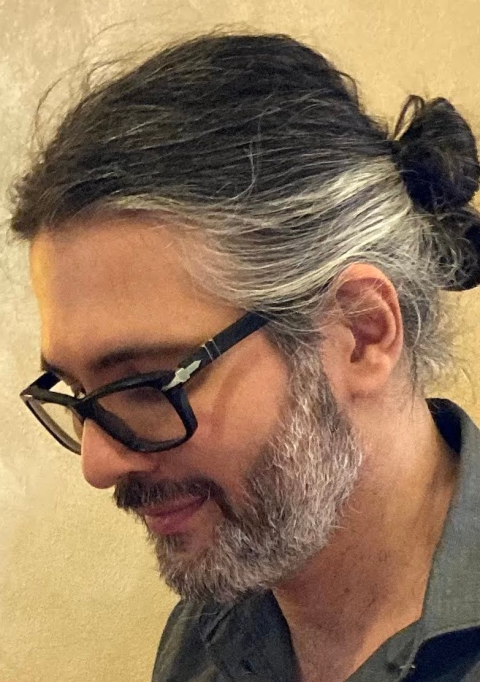}{Giovanni Merlino}{is an Associate Professor at the Department of Engineering, University of Messina, Italy, and a member of several research groups: the \textit{Mobile and Distributed System Lab} (MDSLAB), the \textit{Protocolized Software-defined and Decentralized intelligent Agentic Cyber-Physical Systems} (ProSoDiAC), \textit{the National Interuniversity Consortium for Informatics} (CINI), and head of the IoT laboratory. He holds an International Research Doctorate in Computer and Telecommunications Engineering from the University of Catania. His interests include the Internet of Things for Industry 4.0, fog computing and the Web of Smart Things, cyber-physical systems as software-defined infrastructure, IT/OT convergence, and biomedical applications of deep learning.}

\end{document}